\documentclass[apj]{emulateapj}
\newcount\listnorom
\newcommand\listromanDE{\global\advance \listnorom by 1
{\lowercase\expandafter{(\romannumeral\listnorom)}\ }}
\newcommand\newlistroman{\listnorom=0}

\newcommand{\Eth}{E_\mathrm{th}}
\newcommand{\xx}[1]{\!\times\!10^{#1}}

\newcommand{\FP}{Analytic Precursor Approximation}
\newcommand{\FPshort}{APA}

\newcommand{\Lcell}{L_\mathrm{cell}}
\newcommand{\tstep}{t_\mathrm{step}}

\newcommand{\ElPlasmafreq}{\omega_{pe}}

\newcommand{\protongyrofreq}{\omega_{\mathrm{cp}}}
\newcommand{\MFA}{magnetic field amplification}
\newcommand{\rel}{relativistic}
\newcommand{\NL}{nonlinear}
\newcommand{\CR}{cosmic ray}

\newcommand{\nonrel}{non\-rel\-a\-tiv\-is\-tic}

\newcommand{\mc}{Monte Carlo}
\newcommand{\MC}{Monte Carlo}
\newcommand{\syn}{synchrotron}

\newcommand{\kmps}{km s$^{-1}$}
\newcommand{\pcc}{cm$^{-3}$}
\newcommand{\muG}{$\mu$G}

\newcommand{\SC}{self-consistent}
\newcommand{\SCly}{self-consistently}
\newcommand{\Pcr}{P_\mathrm{cr}}

\newcommand{\Pcrhat}{P_\mathrm{cr}}
\newcommand{\Pth}{P_\mathrm{th}}
\newcommand{\Pthzero}{P_\mathrm{th0}}
\newcommand{\Pwzero}{P_\mathrm{w0}}
\newcommand{\Pwone}{P_\mathrm{w1}}
\newcommand{\Pwtwo}{P_\mathrm{w2}}
\newcommand{\Fwzero}{F_\mathrm{w0}}

\newcommand{\Fwtwo}{F_\mathrm{w2}}
\newcommand{\Ppzero}{P_\mathrm{p0}}

\newcommand{\Pptwo}{P_\mathrm{p2}}
\newcommand{\wpzero}{w_\mathrm{p0}}

\newcommand{\wptwo}{w_\mathrm{p2}}
\newcommand{\Rtot}{r_\mathrm{tot}}
\newcommand{\rtot}{r_\mathrm{tot}}

\newcommand{\rsub}{r_\mathrm{sub}}
\newcommand{\Mazero}{M_\mathrm{A0}}
\newcommand{\Mszero}{M_\mathrm{s0}}
\newcommand{\Msone}{M_\mathrm{s1}}

\newcommand{\vazero}{v_\mathrm{A0}}
\newcommand{\xFEB}{x_\mathrm{FEB}}
\newcommand{\xFP}{x_\mathrm{APA}}
\newcommand{\xtr}{x_\mathrm{tr}}
\newcommand{\fcr}{f_\mathrm{cr}}
\newcommand{\AccEff}{{\cal E}_\mathrm{cr}}

\newcommand{\pmax}{p_\mathrm{max}}
\newcommand{\Emax}{E_\mathrm{max}}

\newcommand{\rgzero}{r_\mathrm{g0}}
\newcommand{\tacc}{\tau_\mathrm{acc}}
\newcommand{\fAlf}{f_\mathrm{Alf}}
\newcommand{\Beff}{B_\mathrm{eff}}

\newcommand{\Befftwo}{B_\mathrm{eff2}}
\newcommand{\Btrend}{B_\mathrm{trend}}
\newcommand{\Qesc}{Q_\mathrm{esc}}
\newcommand{\qesc}{q_\mathrm{esc}}
\newcommand{\taucoll}{\tau_\mathrm{coll}}
\newcommand{\taucomp}{\tau_\mathrm{comp}}
\newcommand{\momentumflux}{\Phi_P}
\newcommand{\upstreammomentumflux}{\Phi_\mathrm{P0}}

\newcommand{\energyflux}{\Phi_E}
\newcommand{\upstreamenergyflux}{\Phi_\mathrm{E0}}

\newcommand{\gammabar}{\bar{\gamma}}
\newcommand{\deltabar}{\bar{\delta}}
\newcommand{\pvector}{{\bf p}}
\newcommand{\vvector}{{\bf v}}

\newcommand{\coldism}{cold ISM ($T_0=10^4$~K)}
\newcommand{\hotism}{hot ISM ($T_0=10^6$~K)}
\newcommand{\coldismNoT}{cold ISM}
\newcommand{\hotismNoT}{hot ISM}

\newcommand{\avemail}{avladim@ncsu.edu}
\newcommand{\abemail}{byk@astro.ioffe.ru}
\newcommand{\deemail}{don\_ellison@ncsu.edu}
\newcommand{\heatpar}{\alpha_H}
\newcommand{\Alf}{Alfv\'{e}n}
\newcommand{\Lbar}{L}

\shorttitle{Turbulence Dissipation in DSA}
\shortauthors{Vladimirov, Bykov, \& Ellison}

\begin{document}

\title{Turbulence Dissipation and Particle Injection in Non-Linear
  Diffusive Shock Acceleration with Magnetic Field Amplification}

\author{Andrey E. Vladimirov} \affil{Physics Department, North Carolina
State University, Box 8202, Raleigh, NC 27695; \avemail}

\author{Andrei M. Bykov} \affil{Department of Theoretical Astrophysics,
Ioffe Physical-Technical Institute, St. Petersburg, Russia; \abemail}

\and

\author{Donald C. Ellison} \affil{Physics Department, North Carolina
State University, Box 8202, Raleigh, NC 27695; \deemail}

\begin{abstract}

The highly amplified magnetic fields suggested by observations of some
supernova remnant (SNR) shells are most likely an intrinsic part of
efficient particle acceleration by shocks.
This strong turbulence, which may result from cosmic ray driven instabilities,
both resonant and non-resonant, in the shock precursor, is certain to
play a critical role in \SC, \NL\ models of strong, \CR\ modified shocks.
Here we present a Monte Carlo model of nonlinear diffusive shock
acceleration (DSA) accounting for magnetic field amplification through
resonant instabilities induced by accelerated particles, and including
the effects of dissipation of
turbulence upstream of a shock and the subsequent precursor plasma
heating.  Feedback effects between the plasma heating due to
turbulence dissipation and particle injection are strong, adding to the
\NL\ nature of efficient DSA.
Describing the turbulence damping in a parameterized way, we reach two
important results: first, for conditions
typical of supernova remnant shocks, even a small
amount of dissipated turbulence energy ($\sim 10\%$) is sufficient to
significantly heat the precursor plasma, and second,
the heating upstream of the shock leads to an increase in the
injection of thermal particles at the subshock by a factor
of several.
In our results, the response of the non-linear shock
structure to the boost in particle injection prevented
the efficiency of particle acceleration and magnetic
field amplification from increasing.
We argue, however, that more advanced (possibly, non-resonant) models
of turbulence generation and dissipation may lead to
a scenario in which particle injection boost due
to turbulence dissipation results in more efficient
acceleration and even stronger amplified magnetic fields than
without the dissipation.

\end{abstract}


\section{Introduction}
Observations of young supernova remnants (SNRs) suggest that strong
collisionless shocks can simultaneously
place a large fraction of the shock ram kinetic energy into \rel\ protons
\citep[e.g.,][]{BE87,MD2001, WarrenEtal2005}
and amplify the ambient turbulent magnetic field by large factors
\citep[e.g.,][]{Cowsik80, RE92, BambaEtal2003, BKV2003, VL2003,
UA2008}.
This coupling of diffusive shock acceleration (DSA) and magnetic field
amplification (MFA) is critically important because the self-generated
magnetic field largely determines the efficiency of DSA, the maximum
particle energy a given shock can produce, and the \syn\ emission from
radiating electrons.

The generation and dissipation of strong MHD turbulence in
collisionless, multi-fluid plasmas is a complex process.
Different \NL\ approaches to the modeling of the large scale
structure of a shock undergoing  efficient cosmic ray acceleration
\citep[e.g.,][]{MV82,AB86,BL2001,AB2006,VEB2006,PLM2006,ZPV2008} have
predicted the presence of strong MHD turbulence in the shock
precursor.
However, an exact modeling of the shock structure in a turbulent medium,
including nonthermal particle injection and acceleration, requires a
nonperturbative, self-consistent description of a multi-component and
multi-scale system including the strong MHD-turbulence dynamics.
While a number of analytic models describing resonant and non-resonant
amplification and damping of magnetic fluctuations have been proposed,
these generally rely on the quasi-linear approximation that the
fluctuations are small compared to the background magnetic field, i.e.,
$\Delta B \ll B_0$ \citep[e.g.,][]{Wentzel74,BT2005,Kulsrud2005}.
No consistent analytic description of magnetic turbulence generation
with $\Delta B \gtrsim B_0$ exists. For these reasons, numerical models
with varying ranges of applicability have been proposed which offer a
compromise between completeness and speed
\citep[e.g.,][]{Bell2004,AB2006,VEB2006,ZPV2008}.

In principle, the problem can be solved completely with few assumptions
and approximations with particle-in-cell (PIC) simulations
\citep[e.g.,][]{Bell2004,Spitkovsky2008,NPS2008}
or, in the assumption that electrons are not dynamically
important, by hybrid models \citep[e.g.,][]{WO1996, Giacalone2004}.
However, modeling the \NL\ generation of \rel\ particles and strong
magnetic turbulence in collisionless shocks is
computationally challenging and PIC simulations will not be able
to fully address this problem in \nonrel\ shocks for some years to come
even though they can provide critical information on the plasma
processes producing injection that can be obtained in no other way. In
Appendix~\ref{estimatesforpic} we outline the requirements that a
PIC simulation must fulfill in order to tackle the problem of efficient
DSA with non-linear MFA in SNR shocks.

In the \mc\ approximation we use here, the plasma interactions are
parameterized allowing us to study coupled \NL\ effects between the
extended shock precursor and the gas subshock. In particular, we
investigate the \NL\ effects caused by upstream plasma heating due to
magnetic field dissipation.

The importance of the dissipation of turbulence in the shock precursor
can be illustrated by the following estimate. Suppose that in a shock
wave of speed $u_0$, the turbulence is generated by the
resonant cosmic ray (CR) streaming instability, so the energy density of
the turbulence, $U_w$, evolves approximately as
$u_0\,dU_w(x)/dx=v_A\,d\Pcr(x)/dx$ \citep[e.g.,][]{BL2001}, where
$\Pcr(x)$ is the CR pressure at position $x$ and $v_A$ is the
\Alf\ speed.
Ignoring all non-linear effects, the turbulence energy density at
the shock positioned at $x=0$ is $U_w(0)=\rho_0 u_0 \vazero \cdot
\Pcr(0)/(\rho_0 u_0^2)$.
The ratio $\AccEff=\Pcr(0)/(\rho_0 u_0^2)$ characterizes the
efficiency of acceleration and is typically assumed to be on the
order of ten percent or more.
In the above, $\rho$ is the fluid density and the subscript ``0''
always indicates far upstream values.
Suppose a fraction, $\heatpar$, of this energy goes into heating of the
thermal gas in the shock precursor so the energy density of the thermal
plasma increases by $\Delta U_{H}(0) = \heatpar U_w(0)$ at $x=0$.
Comparing $\Delta U_{H}(0)$ with the internal energy density of the far
upstream plasma, $\epsilon_0$, we find
\begin{equation}
\label{upsilonequation} \eta_H\ = \frac{\Delta U_H(0)}{\epsilon_0}
\approx
   \heatpar \AccEff \frac{M_s^2}{M_A}
\ ,
\end{equation}
where $M_s$ is the sonic, and $M_A$ the \Alf, Mach number. If
$\eta_H$ is large, the thermal gas in the shock precursor is
strongly heated and this influences the subshock strength and the
particle injection efficiency. In a non-linearly modified shock, a
change in the injection efficiency causes the whole shock structure
to change.
For typical supernova remnant (SNR) parameters (e.g., $u_0 \sim
3000$ \kmps, $T_0 \sim 10^4$ K, $n_0 \sim 0.3$ \pcc, and $B_0 \sim
3$ \muG),
the ratio $M_s^2/M_A \approx 250$, and values of $\heatpar$ as low
as a few to ten percent may be important.

Because existing analytical descriptions of MHD wave damping
rely on the quasi-linear approximation $\Delta B \ll B_0$, which is
inapplicable for strong turbulence,
and because an exact description of this process
in the framework of non-linear DSA is currently impossible
(see Appendix~\ref{estimatesforpic}), we propose a
parameterization of the turbulence damping rate.
In doing this, we are pursuing two goals.
First, we make some predictions connecting cosmic
ray spectra, turbulent magnetic fields
and plasma temperatures, which, in principle, can be tested
against high resolution X-ray observations in order to estimate the
heating of the thermal gas by turbulence dissipation.
And second, once heating is included in our simulation in a
parameterized fashion, we will be ready to implement more realistic
models of turbulence generation and dissipation as they are developed.

Our \mc\ simulation can be briefly summarized as follows \citep[see][ and
references therein for more complete details]{EJR90,JE91,VEB2006}.
We describe particle transport in a plane shock by Bohm diffusion in a
plasma flowing in the $x$-direction with speed $u(x)$.
Particles move in small time steps as their local plasma frame momenta
are `scattered' at each step in a random walk process on a sphere in
momentum space.
Some shock heated thermal particles are injected into the
acceleration process when their history of random scatterings in the
downstream region takes them back upstream.  These particles gain
energy and some continue to be accelerated in the first-order Fermi
mechanism. This form of injection is generally called `thermal
leakage' and was first used in the context of DSA in \citet{EJE1981}
\citep[see also][]{Ellison82}.

The magnetic field determining the random walk properties
through the diffusion coefficient
is the `seed' (interstellar) magnetic field after it has been amplified
by a large factor by the CR streaming instability and compressed and
advected downstream with the flow.  The streaming instability in the
non-linear regime ($\Delta B \gg B_0$) is described by the traditional
quasi-linear equations, where the instability driving term is the CR
pressure gradient.
These quasi-linear equations are extrapolated into the non-linear regime
in a parameterized fashion for lack of a more complete analytic
description. The magnetic turbulence generated by the instability is
assumed to dissipate at a rate proportional to the turbulence generation
rate, and the dissipated energy is pumped directly into the thermal
particle pool.  An iterative scheme is employed to ensure the
conservation of mass, momentum, and energy fluxes, thus producing a
self-consistent solution of a steady-state, plane shock, with particle
injection and acceleration coupled to the bulk plasma flow
modification and to the magnetic field amplification and damping.

Our results show that even a small rate of turbulence dissipation
can significantly increase the precursor temperature and that this, in
turn, can increase the rate of injection of thermal particles. The
\NL\ feedback of these changes on the shock structure, however, tend
to cancel so that the spectrum of high energy particles is only
modestly affected.

\section{Model}
The \mc\ simulation used here contains all of the elements of the code
used in \citet{VEB2006}, and previously by Ellison and co-workers, i.e.,
it iteratively determines a \SC, steady-state shock solution with
resonant \MFA. The present code, however, has been completely re-written
and optimized for the problem of DSA with MFA in nonrelativistic, plane
shocks. Besides including dissipation, the new code has been written for
parallel processing and can model acceleration, in a reasonable time, to
PeV energies\footnote{The ability of the Monte Carlo code to
accelerate particles to PeV energies was shown in \citet{EV2008}. The
results we are interested in here do not require such high energies and
we use a smaller dynamic range (see the note on dynamic range at the end
of Section~\ref{nlresults}).}.
The
particle transport is briefly described in Section
\ref{particletransport}, the amplification and dissipation of turbulence
are described in Section~\ref{amplificationanddamping}, the effects of
turbulence damping on the thermal plasma are described in
Section~\ref{plasmaheating}, and the iterative procedure used to reach a
self-consistent solution is given in Section~\ref{iterativeprocedure}.

\subsection{Particle Transport and Injection in the Monte Carlo Code}
\label{particletransport}
Given a flow speed profile $u(x)$ and a diffusion coefficient $D(x,p)$
(which are obtained as explained below), the Monte Carlo code generates
a thermal distribution of particles far upstream (or close to the shock,
as described in Section \ref{plasmaheating} and Appendix~\ref{fastpush}),
and propagates them by small time steps,
$\delta t$, scattering their momenta with the `pitch-angle scattering'
scheme described in detail in \citet{EJR90}.
Since we assume strong turbulence ($\Delta B \gg B_0$), the concept of
`pitch-angle' as an angle measured with respect to the average magnetic
field direction loses it meaning. Instead, we define the `pitch-angle'
as the angle that a particle's plasma frame momentum makes with the
direction of the bulk flow.  Each scattering is elastic and
isotropic in the local plasma frame, which moves at speed $u(x)$ with
respect to the viscous subshock located at $x=0$, and the angle of
scattering is random, but its maximal value is determined by $\delta t$
and by $D(x,p)$ \citep[see][]{EJR90}. When particles cross the subshock,
or move in any compressive flow, the elastic scattering in the plasma
frame makes them gain energy in the shock frame (see also
Section~\ref{mcadiabatic}).

Injection of particles into the acceleration process occurs in the Monte
Carlo simulation when a formerly thermal particle first crosses the
viscous subshock backwards, i.e., going against the flow.
The number of particles that do this, and the energy they gain, are
determined only by the random particle histories; no parameterization of
the injection process is made other than the assumption of the diffusion
coefficient value at various particle energies.\footnote{As in previous
implementations of our \mc\ model, the subshock is assumed to be
transparent to all particles, including thermal ones, and possibly
important plasma effects such as a cross-shock potential or large
amplitude magnetic structures in the subshock layer are ignored.}

Since microscopic plasma modeling of particle injection processes in
non-relativistic subshocks with an appropriate 3-D PIC code is not
yet available, all the models dealing with particle acceleration by
shocks must assume  some prescription of the injection process. We
favor the present model for two reasons. First, our injection rate
is set by the scattering prescription and doesn't require any
additional assumptions. Second and even more important, our
model was shown to agree well with spacecraft observations of the
Earth's bow shock \citep[][]{EM87, EMP90}, interplanetary shocks
\citep[][]{BOEF97}, and with 1-D hybrid PIC simulations
\citep[][]{EGBS93}. We are aware of alternative models of particle
injection, such as that used by \citet{BGV2005}, in which the shock is assumed
transparent only to particles exceeding the thermal gyroradius by a
certain factor. It is possible to simulate the injection with the
\citet{BGV2005} recipe in the Monte Carlo code, which was done by
\citet{EBG2005}, but it should be noted that an injection threshold
may be inconsistent with the bow shock observations reported in
\citet{EMP90}. Our thermal injection model is also simultaneously
consistent with the bow shock helium and CNO observations with no
additional parameters or assumptions. A comparative analysis of the
different injection recipes is beyond the scope of our present study.

A peculiarity of our approach is that in order to separate the CR
particles from the thermal ones we use their history, and not their
energy.  By our definition, a thermal particle is one that we had
introduced into the simulation upstream with a random thermal energy and
that may have crossed the subshock going downstream, but has never crossed
it back. Once a particle crosses the subshock (the coordinate $x=0$, to
be more precise) in the upstream direction, it by our definition
is injected and becomes a CR particle
(see also Appendix~\ref{special_th_vs_cr}).

As particles are propagated and scattered, their contributions to fluxes
of mass, momentum, and energy are calculated at select positions.  We
also calculate the pressure produced by the thermal particles $\Pth(x)$
and the spectrum of pressure of the CR particles $\Pcrhat(x,p)$.
(see details in Appendix~\ref{fastpush}). The latter is
related to the total CR pressure $\Pcr(x)$ as
\begin{equation}
\Pcr(x) = \int\limits_0^{\infty}{\Pcrhat(x,p)dp}
\ ,\end{equation}
and $\Pcrhat(x,p)$ is then used to calculate the magnetic field
amplification and dissipation, as described in Section
\ref{amplificationanddamping}.\footnote{We note that notation has
changed slightly from that used in \citet{VEB2006}. The subscript `tot'
indicating integral quantities has been eliminated, and these quantities
are now denoted as $\Pcr(x)$, $L(x)$, etc. The quantities representing
spectral densities of pressure, flux, etc., are denoted with the same
letters, but a different set of arguments: $\Pcr(x,p)$, $L(x,k)$. To
avoid ambiguity, explicit definitions relating the spectral densities to
the integral densities are provided throughout the text. We
also dropped the bar above the dissipation rate term $L(x)$
for simplicity.}

We assume in this paper that the acceleration is size-limited, and model
the finite size of the shock with a free escape boundary (FEB), located
at position $\xFEB<0$ far upstream of the shock. All CR particles
crossing the boundary escape freely from the system. The maximal
particle momentum, $\pmax$, is thus determined when
the upstream diffusion length becomes comparable to $|\xFEB|$. For a
spherical SNR blast wave, $|\xFEB|$ is on the order of the radius of the
remnant.

\subsection{Turbulence Amplification and Dissipation}

\label{amplificationanddamping}

We assume that far upstream there exists a uniform magnetic field,
$B_0$, parallel to the flow direction which is perturbed by transverse
\Alf ic fluctuations with a power law energy spectrum. It is further
assumed that these fluctuations produce Bohm diffusion for particles
of all energies.
Closer to the shock, where a population of accelerated particles
drifting upstream is present, this seed turbulence is amplified by the
CR streaming instability and additionally strengthened by the plasma
compression. Once amplification begins, the spectrum of turbulence is no
longer restricted to any particular form\footnote{We assumed
$U_{\pm}\propto k^{-1}$ for the seed turbulence.  The choice of the seed
turbulence spectrum does not significantly affect our results
for two reasons. First, the diffusion coefficient assumed here only
depends on the total power in the turbulence and is insensitive to the
shape of the spectrum, and second, the rapidly growing fluctuations due
to the streaming instability quickly overpower the seed
spectrum. Despite our using the Bohm diffusion coefficient, we still
keep track of the turbulence spectrum in the simulation since this
will be used in future work where the diffusion coefficient is
determined self-consistently from the wave spectrum.}.
Assuming that the turbulence is described by the
quantities $U_-(x,k)$ and $U_+(x,k)$ ($k$ is the wavenumber of
turbulence harmonics, $U_-$ and $U_+$ are the spectra of energy
density of structures propagating in the upstream and the downstream
directions with respect to the thermal plasma, respectively), we model
the evolution of the turbulence, as it is being advected with the plasma
and amplified, with the following equations:
\begin{equation}
\label{ampeq}
E_{\pm}[U] = (1-\heatpar)G_{\pm}[U] + I_{\pm}[U].
\end{equation}
Here, for readability, we abbreviated as $E$ the evolution operator,
as $G$ the growth operator and as $I$ the wave-wave interactions
operator, acting on the spectrum of turbulence energy density
$U=\{U_-(x,k), U_+(x,k)\}$. These quantities are defined as follows:
\begin{eqnarray}
\nonumber
  E_{\pm}[U] & = &
     \left(u \pm V_G\right)\frac{\partial}{\partial x} U_{\pm} + \qquad \\
     && \qquad U_{\pm}\frac{d}{dx}\left(\frac32 u \pm V_G\right), \\
\nonumber
  G_{\pm}[U] & = &
     \mp\frac{U_\pm}{U_+ + U_-} V_G \times \qquad \\
     && \qquad \frac{\partial \Pcrhat(x,p)}{\partial x}
     \left| \frac{dp}{dk} \right|, \\
  I_{\pm}[U] & = &
     {\pm} \frac{V_G}{\rgzero}\left(U_- - U_+\right).
\end{eqnarray}
Here and throughout the paper, $\rgzero=m_p u_0 c / (eB_0)$ -- the
gyroradius of a particle with a plasma frama speed equal to $u_0$
in the far upstream magnetic field $B_0$.
The parameter $\heatpar$ describes the turbulence dissipation rate, and
for $\heatpar=0$, equation~(\ref{ampeq}) is exactly
what was used in \citet{VEB2006}, except there it was
assumed that $V_G\ll u(x)$.  Keeping $V_G$ relative to $u(x)$ results in
a slightly greater amplification of magnetic field.
In this system $u=u(x)$ is the flow speed and
$V_G=V_G(x)$ is the parameter defining the turbulence growth rate
and the wave speed\footnote{As explained in \citet{VEB2006},
in the quasilinear case, $\Delta B \ll B_0$,
the wave speed and the speed determining turbulence growth rate are
both equal to the \Alf\ speed,
$V_G (x) = v_A = B_0 / \sqrt{4 \pi \rho(x)}$.
In the case of strong turbulence, $\Delta B \gtrsim B_0$,
we hypothesise that the resonant streaming instability can
still be described by equations~(\ref{ampeq}) with $V_G$
being a free parameter ranging from $B_0 / \sqrt{4 \pi \rho(x)}$
to $\Beff / \sqrt{4 \pi \rho(x)}$.}.

For simplicity, we assume $V_G (x) = B_0 /\sqrt{4 \pi \rho(x)}$,
in the present work, corresponding to $\fAlf=0$ in \citet{VEB2006},
and emphasize that the use of equation~(\ref{ampeq}) to
describe the streaming instability when $\Delta B \gtrsim B_0$ is only a
parameterization.

The growth operator $G$, which describes the turbulence amplification by
the CR streaming instability, is proportional to the gradient of
$\Pcr(x,p)$, the latter being the spectrum of pressure of the CR
particles driving the instability.  $\Pcr(x,p)$ is
computed in the Monte Carlo simulation from the trajectories of
particles, and the momentum $p$ at which it is taken in (\ref{ampeq}) is
the momentum resonant with the turbulent structures with wavenumber
$k$. The resonance condition assumed is
\begin{equation}
k \frac{c p}{e B_0} = 1.
\end{equation}

The parameter  $\heatpar$ enters the
equations of turbulence evolution (\ref{ampeq}) through the factor
$(1-\heatpar)$, by which the term $G$ describing the magnetic field
amplification by the CR streaming is reduced (it is assumed that $0 \leq
\heatpar \leq 1$). By writing this, we assumed simply that at all
wavelengths only a fraction $(1 -\heatpar)$ of the
instability growth rate goes into the magnetic turbulence, and the
remaining fraction $\heatpar$ is lost in the dissipation process.
The factor $(1-\heatpar)$ in (\ref{ampeq}) can also be understood in the
following way: \citet{VEB2006} derived their equations~(11) and (12) from
equation (3) by assuming that the loss term $L=0$ ($L$ here is $\bar{L}$ in
their notation), but now we
are assuming that
\begin{equation}
\label{dampingparameterization_spc}
  L=- \heatpar \cdot v_\mathrm{a, x}\frac{\partial \Pcr}{\partial x} .
\end{equation}
For $\heatpar=0$ no dissipation occurs, and the
CR streaming instability pumps the energy of the accelerated particles
into the magnetic turbulence, amplifying the effective magnetic field
most efficiently.  For $\heatpar=1$
the additional turbulence energy produced by the instability is assumed
to immediately dissipate, and the scattering in the shock is assumed to
be provided only by the seed turbulence slightly increased by the plasma
compression.

The energy dissipation rate at all wavelengths is then
\begin{equation}
\label{dampingparameterization}
L(x)=\int\limits_0^{\infty} L(x,k)\: dk = \heatpar \cdot V_G \frac{d
\Pcr(x)}{dx},
\end{equation}
and we assume that all this energy goes directly into heating
thermal particles. The
modeling of the thermal plasma heating is covered in detail in Section
\ref{plasmaheating}.

When equation~(\ref{ampeq}) is solved, the resulting
$U_{\pm}(x,k)$ are used to calculate the amplified effective magnetic
field
\begin{eqnarray}
\nonumber
\Beff(x) &=& \left[ 4 \pi \int\limits_0^{\infty} U_-(x,k)\:dk +\qquad \right. \\
\label{beffdef}
&& \left. 4\pi \int\limits_0^{\infty} U_+(x,k) \: dk \right]^{1/2},
\end{eqnarray}
the turbulence pressure
\begin{equation}
\label{pwdef}
P_w(x) = \frac{\Beff^2(x)}{8 \pi}
\end{equation}
and the turbulence energy flux
\begin{eqnarray}
\nonumber
F_w(x) &=& \left( \frac32 u - V_G \right) \int\limits_0^{\infty}U_-(x,k)\:dk+\qquad\\
\label{fwdef}
&&\left( \frac32 u + V_G \right) \int\limits_0^{\infty}U_+(x,k)\:dk
\end{eqnarray}
(see equations (15), (18) and (19) in \citet{VEB2006}), which are then
used in the derivation of a self-consistent solution, as discussed in
Section~\ref{iterativeprocedure} below. In the present paper we do
not neglect $V_G$ in the sum with $u$, which results in a slightly lower
compression ratio $\rtot$ than in the case $3/2 u \pm V_G \approx 3/2
u$, that was adopted in \citet{VEB2006}. Note that the factor
$3/2$ is valid for \Alf\ wave-like modes, which is implicitly assumed 
by our using the system of equations (\ref{ampeq}).

As mentioned by \citet{CBAV2008}, effects from the transmission and
reflection of \Alf\ waves at the subshock could be important, and
accounting for these effects may further lower the compression ratio
$\rtot$.
We do not account for reflection and linear transformations of \Alf\
waves to other MHD modes (magnetosonic, entropy, etc.) at the subshock
(see McKenzie and Westphal 1970) in the present paper, because a correct
description of these effects on the subshock in highly turbulent media
must contain simultaneously some other comparable effects.
Indeed, MHD waves interacting with a shock produce stochastic ripples in
the shock surface and these ripples produce an effective broadening of
the shock spatial structure determined by the turbulence spectrum
\citep[see][]{Bykov1982}.
Moreover, as we argue in Appendix~\ref{special_th_vs_cr}, accounting for
suprathermal particles modifies the standard Rankine-Hugoniot relations
at the subshock and these effects could make it impossible to identify
the subshock as a plane discontinuity, as assumed in all existing
semi-analytic models.
Clearly, these phenomena are important and require further investigation
but they are beyond the scope of our simplified description of
MFA. Nevertheless, we believe our predictions regarding the turbulence
dissipation in the shock precursor are qualitatively correct.

\subsection{Heating of Thermal Plasma}
\label{plasmaheating}
Repeating the derivation of equation (9) in \citet{MV82}, one
obtains for a steady-state shock:
\begin{equation}
\label{pressuregrowth}
\frac{ u\rho^{\gamma}}{\gamma-1}
\frac{d}{dx}\left(\Pth \rho^{-\gamma} \right) = \Lbar(x)
\ .
\end{equation}
Here and elsewhere, $\rho=\rho(x)$, $u=u(x)$, and $\Pth=\Pth(x)$.
The quantity $\Lbar(x)$ is the dissipation rate defined in
(\ref{dampingparameterization}), and the ratio of specific
heats of an ideal nonrelativistic gas is $\gamma=5/3$.
For $\Lbar(x)=0$, equation (\ref{pressuregrowth})
reduces to the adiabatic heating law,
$\Pth \sim \rho^{\gamma}$ and, for a non-zero $\Lbar(x)$, it describes
the heating of the thermal plasma in the shock precursor due to the
dissipation of magnetic turbulence.
The fluid description of heating given by
equation~(\ref{pressuregrowth}), while it doesn't include details of
individual particle scattering, can be used in the \mc\ simulation to
replace particle scattering and
determine heating in the shock precursor.  This merging of analytic and
\mc\ techniques, or \FP\ (\FPshort), is described in detail
in Appendix \ref{fastpush}.

The \FPshort\ only affects our treatment of thermal
(i.e., not injected)
particles in the precursor and involves two steps. The first is to
introduce thermal
particles into the simulation, not far upstream as we did before, but at
some position $\xFP < 0$ close to the subshock and with a temperature
equal to what Eq.~(\ref{pressuregrowth}) and the ideal gas law
suggest:
\begin{equation}
\label{idealgaslaw}
T(x)=\frac{\Pth(x)}{k_B n_0 (u_0/u(x))},
\end{equation}
where all quantities are taken at $x=\xFP$ and $\Pth$ is calculated from
(\ref{pressuregrowth}) ($n_0$ and $u_0$ are the far upstream number
density and shock speed, and $k_B$ is
Boltzmann's constant).  The second step is to calculate the
thermal gas pressure throughout the precursor (at $x<\xFP$) using
(\ref{pressuregrowth}) instead of tracing particle motions.

The main effects of turbulence dissipation in our model are:
\newlistroman\listromanDE a decrease in the value of the amplified field
$\Beff(x)$, which determines the diffusion coefficient, $D(x,p)$;
\listromanDE an increase in the temperature of particles just upstream of
the subshock, which influences the injection of particles into the
acceleration process, and
\listromanDE
an increase in the thermal particle pressure $\Pth(x<0)$,
and a decrease in the turbulence pressure $P_w(x)$,
which enter the conservation equations
described in Section \ref{iterativeprocedure}.
Since all of these processes are coupled, a change in dissipation
influences the overall structure of the shock.

\subsection{Closure of the Model}
\label{iterativeprocedure}
Typically, we begin our simulation by propagating particles, with a
pre-set diffusion coefficient $D(x,p)$, in an unmodified shock, where the
flow speed jumps discontinuously from $u_0$ to $u_2$, that is,
$u(x)=u_0$ for $x<0$ and $u(x)=u_2$ for $x>0$.\footnote{Everywhere in
the text, unless otherwise noted,
the subscript `0' indicates a far upstream value, `1'
indicates a value just upstream of the subshock, and `2' indicates a
downstream value. For example, $u_0=u(x=-\infty)$, $u_1=u(\xtr)$, and
$u_2=u(x>0)$. See Appendix~\ref{mcadiabatic} for definition of $\xtr$.}
This allows us to calculate the various fluxes and other quantities,
such as the CR pressure spectrum $\Pcrhat(x,p)$, at any position $x$.
The latter is used to solve Eq.~(\ref{ampeq}), which yields the
turbulence spectra $U_{\pm}(x,k)$ and, subsequently, the amplified
effective field $\Beff(x)$ and the pressure of the magnetic
turbulence $P_w(x)$. The spectrum $\Pcrhat(x,p)$ also provides
the turbulence dissipation rate (\ref{dampingparameterization})
and the resulting pressure of the turbulence-heated gas
(\ref{pressuregrowth}).

The equations for the conservation of mass and momentum fluxes are:
\begin{equation}
\label{fluxmass}
\rho(x) u(x) = \rho_0 u_0
\end{equation}
($\rho$ and $u$ are the mass density and the flow speed) and
\begin{equation}
\label{fluxmomentum}
\momentumflux(x) =  \upstreammomentumflux,
\end{equation}
where
$\momentumflux(x)$ is the flux of the $x$-component of momentum in the
$x$-direction including the contributions from particles and turbulence,
and $\upstreammomentumflux$ is the far upstream value of momentum flux, i.e.,
\begin{equation}
\upstreammomentumflux  =
   \rho_0 u_0^2 + \Pthzero + \Pwzero
\ .
\end{equation}
The quantity $\momentumflux$ is defined as
\begin{equation}
  \label{mfasmoment}
  \momentumflux(x) = \int p_x v_x f(x, \pvector) d^3p + P_w(x),
\end{equation}
(here $p_x$ and $v_x$ are the $x$-components of momentum and
velocity of particles, and $f(\pvector)$ is their distribution
function, all measured in the shock frame),
and in the simulation it is calculated by summing the contributions
of the particles crossing a grid location and adding the turbulence
pressure $P_w$ defined in (\ref{pwdef}). See details of this computation in
Appendix~\ref{fastpush}.

Initially, in our simulation, the shock doesn't have a \SC\
structure because we start with an unmodified shock and
$\momentumflux(x)$ is overestimated at all locations where
accelerated particles are present.
The next step is to use the calculated macroscopic quantities to find a
new $u(x)$ that reduces the mismatch between the local momentum
flux and the far upstream value of it $\upstreammomentumflux$
for $x<0$. We do this by calculating
\begin{eqnarray}
\label{iteration}
  u'(x)=u(x) + s \cdot \frac{\momentumflux(x) - \upstreammomentumflux}{\rho_0 u_0},
\end{eqnarray}
where $u'(x)$ is the predicted flow speed for the
next iteration, and $s$ is a small positive number (typically around
0.1), characterizing the pace of the iterative procedure. At this
point we also refine our
estimate for the particle diffusion coefficient which, as in
\citet{VEB2006}, is assumed to be Bohm diffusion such that the particle
mean free path is equal to its gyroradius in the effective, amplified
field $\Beff(x)$:
\begin{equation}
\label{bohm}
D(x,p)=\frac{v\lambda}{3} = \frac{vcp}{3e\Beff(x)},
\end{equation}
where $\Beff(x)$ is defined in (\ref{beffdef}).

The predicted $u(x)$ and $D(x,p)$ are then used in a new iteration where
particles are injected and propagated. The calculated CR pressure, momentum
flux, etc. are then used to refine the guesses for $u(x)$ and $D(x,p)$
for the next iteration, and so on. This procedure is continued until all
quantities converge.

In order to conserve momentum and energy, the compression ratio,
$\rtot=u_0/u_2$, must be determined self-consistently with the
shock structure. To determine $\rtot$ we use the condition of the
conservation of energy flux given by:
\begin{equation}
\label{fluxenergy}
\energyflux(x) + \Qesc(x) = \upstreamenergyflux,
\end{equation}
where $\energyflux(x)$ is the energy flux of particles and turbulence in
the $x$-direction, $\Qesc$ is the energy flux of escaping particles
at the FEB,\footnote{Particle escape at an upstream FEB also
causes the mass and momentum fluxes to change but these changes are
negligible as long as $u_0 \ll c$ \citep[][]{Ellison85}.} and the far
upstream value of the energy flux is
\begin{equation}
\upstreamenergyflux = \frac{1}{2} \rho_0 u_0^3 +
\frac{\gamma}{\gamma-1}\Pthzero u_0  + \Fwzero.
\end{equation}
The quantity $\energyflux(x)$ is defined as
\begin{equation}
\label{efasmoment}
\energyflux(x) = \int K v_x f(x,\pvector) d^3 p + F_w(x),
\end{equation}
$K$ being the kinetic energy of a particle with momentum $p$ measured in
the shock frame, and $F_w$ is the energy flux of the turbulence
defined in (\ref{fwdef}). The details of calculating
$\energyflux(x)$ in the simulation are given
in Appendix~\ref{fastpush}, and the explanation of
how $\energyflux(x)$ is used in an iterative procedure converging to a
consistent $\rtot$ is given in Appendix \ref{compressionratio}.

\section{Results}
\subsection{Particle Injection in Unmodified Shocks (Subshock Modeling)}
\label{injectionvsdissipation}

In order to isolate the effects of plasma heating on particle injection,
we first show results for unmodified shocks, i.e., $u(x<0)=u_0$ and
$u(x>0)=u_0/\rtot$, with fixed $\rtot$. In these models particle
acceleration, \MFA\ and turbulence damping are included consistently
with each other, but we do not obtain fully \SC\ solutions
conserving momentum and energy, since this requires the shock
to be smoothed, while we intentionally fix $u(x)$.

%\clearpage
\begin{figure}
\epsscale{1.0}
\plotone{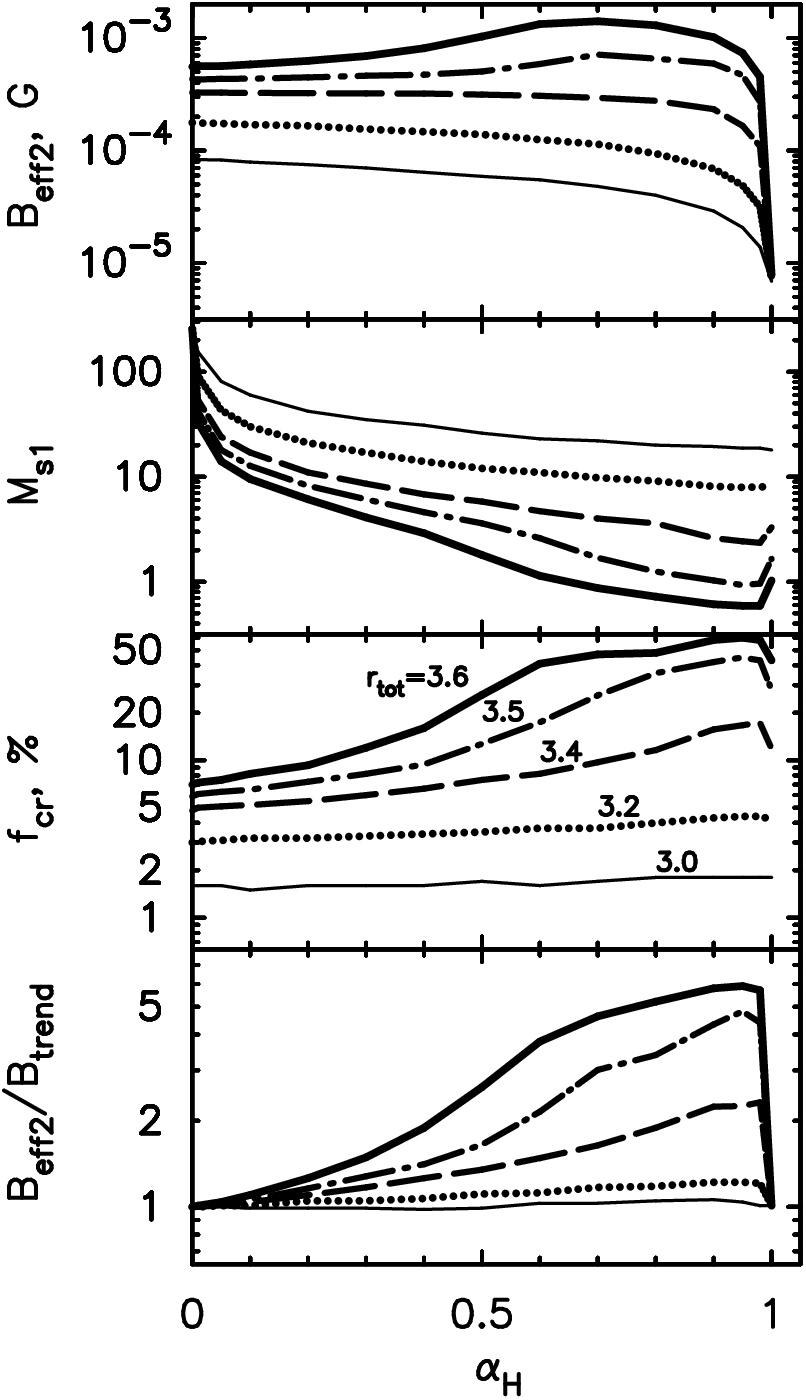}
\caption{Dependence of magnetic field amplification and particle
  injection on the rate of magnetic turbulence dissipation in unmodified
  shocks.  Results for shocks with compression ratios $\rtot$=3.0, 3.2,
  3.4, 3.5 and 3.6 are represented by the thin solid, dotted, dashed,
  dash-dotted and thick solid lines, respectively. The $x$-axis variable
  is the turbulence dissipation rate $\heatpar$ (constant throughout a
  shock), and the plotted quantities are the amplified downstream
  effective magnetic field $\Befftwo$, the subshock sonic Mach number
  $\Msone$, the fraction of simulation particles
  injected into the acceleration process $\fcr$, and the ratio of
  amplified field to $\Btrend$ downstream from the shock.
  \label{fig_unmod_inj}}
\end{figure}
%\clearpage

In Fig.~\ref{fig_unmod_inj} we show results where the compression ratio is
varied between $\Rtot=3$ and $3.6$ as indicated. In all models, $u_0 =
3000$ \kmps, $T_0 = 10^4$ K, $n_0 = 0.3$\,\pcc\ and $B_0 = 3$\,\muG\
(the corresponding sonic and \Alf\ Mach numbers are $\Mszero \approx
\Mazero \approx 250$). The FEB was set at $\xFEB = -3\cdot 10^4\:
\rgzero$ (our spatial scale unit $\rgzero=m_p u_0 c / (eB_0)$),
and for each $\Rtot$ we obtained results for different values
of $\heatpar$ between 0 and 1. The values plotted in the
top three panels of Fig.~\ref{fig_unmod_inj}
are the amplified magnetic field downstream, $\Befftwo$, the Mach number
right before the shock, $\Msone$ (this is not equal to $\Mszero$ because
of the plasma heating due to turbulence dissipation), and the fraction
of thermal particles in the simulation that crossed the shock in the
upstream direction at least once (i.e., got injected), $\fcr$. The bottom
panel shows the ratio of the calculated downstream effective magnetic
field $\Befftwo$ to trend values $\Btrend(\heatpar)$;
what is meant by ``trend'' is explained below.

Looking at the curve for $\Befftwo$ in the $\rtot=3.0$ model, one sees
an easy to explain behavior: as the magnetic turbulence dissipation rate
increases, the value of the amplified magnetic field decreases, going
down to $B_0 \rtot^{3/4}$ (the upstream field compressed at the shock)
for $\heatpar=1$. Increasing $\heatpar$ simply causes more energy to be
removed from magnetic turbulence and put into thermal particles, thus
decreasing the value of $\Befftwo$.\footnote{We use this result, with
well understood behavior, to test the implementation of turbulence
dissipation in our simulation.}
In our model, the amount of dissipated turbulence energy scales linearly
with $\heatpar$. Therefore, the trend of $\Befftwo$ changing with
$\heatpar$ under the assumption that the total efficiency of the
streaming instability is unchanged, but the energy is channelled
from magnetic turbulence to thermal particles, can be described as
\begin{eqnarray}
  \label{btrend}
  \nonumber
  \Btrend^2(\heatpar) &=&
     \left(B_0 \rtot^{3/4}\right)^2 + (1-\heatpar) \times \qquad\quad\\
  \label{btwotrend}
  &&
  \left[ \left. \Befftwo^2 \Big.\right|_{\heatpar=0}
    - \left(B_0 \rtot^{3/4}\right)^2 \right],
\end{eqnarray}
where the first term on the right hand side is the $B_0$
compressed at the shock \citep[the compression factor $\rtot^{3/4}$
is explained by equation (13) in][]{VEB2006}, and the
second term is proportional to the amount of the magnetic
turbulence energy density generated by the instability for $\heatpar=0$,
reduced by the factor $(1-\heatpar)$.  Neglecting $B_0$,
Eq.~(\ref{btwotrend}) predicts $\Btrend \propto
\sqrt{1-\heatpar}$. The comparison of $\Befftwo$ with
$\Btrend(\heatpar)$ from (\ref{btwotrend}) is shown in the bottom panel
of Fig.~\ref{fig_unmod_inj}. It is clear that the
above trend matches the calculations very well for $\rtot=3.0$.
One also can see that the sonic Mach
number in this simulation decreased from the upstream value of $250$ to
approximately $20$, and that the fraction of injected particles remained
almost constant as $\heatpar$ was varied.

The curves for $\rtot=3.2$ demonstrate the same behavior, and the shape
of the $\Befftwo$ curve is similar to the one for $\rtot=3.0$, with the
higher compression ratio producing a higher value of the amplified
magnetic field due to a greater number of particles getting
injected. The calculated $\Befftwo$ deviates from the trend
$\Btrend(\heatpar)$ more than in the $\rtot=3.0$ case, and this
deviation marks the emergence of an effect that becomes more
pronounced as $\rtot$increases.

The plots for $\rtot \gtrsim 3.4$ present a
qualitatively different behavior from those with $\rtot \lesssim
3.2$. The downstream magnetic field $\Befftwo$ does
decrease with increasing $\heatpar$, but not as rapidly as in the
previous two cases, and there is a switching point at $\heatpar \approx
0.95$ in the curves for $\Msone$ and $\fcr$.  The bottom panel of
Fig.~\ref{fig_unmod_inj} shows a deviation of $\Befftwo$ from the trend
(\ref{btwotrend}) by a large factor in the $\rtot = 3.4$ case.  This
effect becomes even more dramatic for $\rtot=3.5$ and $\rtot=3.6$ where
$\Befftwo$, contrary to expectations, increases with $\heatpar$ before
$\heatpar\rightarrow 1$.  The fact that the final energy in turbulence
can increase as more energy is transferred from the turbulence to heat
indicates the \NL\ behavior of the system and shows how sensitive the
acceleration is to precursor heating.

Indeed, if the upstream plasma is not heated sufficiently (as in the
$\rtot < 3.4$ cases), then the thermal particles first reaching the
shock are cold ($T_1 \sim T_0$), and just upstream of the subshock
the sonic Mach number $\Msone$ has a large value (much greater than $10$
for any $\heatpar$).  As these thermal particles cross the shock, their
momenta in the downstream plasma frame lie in a very narrow cone opening
in the downstream direction, with the opening angle around $\theta \sim
\Msone^{-1}$, making it equally difficult for most particles to turn
around, cross the subshock backwards and get injected into the
acceleration process.  As long as $\theta$ is small (or $\Msone$ is
large enough), the number of injected particles is insensitive to the
exact opening angle and $\fcr$ stays relatively constant, as seen in the
third panel of Fig.~\ref{fig_unmod_inj} for the $\rtot=3.0$ case.

The injection rate increases with $\Rtot$ and for $\Rtot=3.6$, even with
$\heatpar$ as low as 0.1, the CR-generated turbulence heats the thermal
plasma through dissipation enough to lower $\Msone$ to $\sim 10$.
Increasing $\heatpar$ further lowers $\Msone$ even more, quickly
increasing the probability that a downstream thermal particle will
return upstream, thus boosting the injection rate ($\fcr$ rapidly goes
up with $\heatpar$ in the $\rtot \geq 3.4$ cases). It turns out that
this effect overcomes the reduction of $\Befftwo$ due to damping, and
$\Befftwo$ starts increasing with $\heatpar$.

For any $\rtot$, at some high enough value of $\heatpar$ (near $0.90$),
the decrease in magnetic turbulence due to dissipation dominates the
increase in injection and the magnetic field drops. As $\heatpar
\rightarrow 1$, and $\Befftwo$ becomes small enough, the efficiency of
particle acceleration reduces sufficiently to cut down the total energy
put into magnetic turbulence and the corresponding fraction of this energy
dissipated into the thermal particles. This causes $\Msone$ to turn up
and $\fcr$ to turn down at about $\heatpar=0.95$,
as shown in the Fig.~\ref{fig_unmod_inj}.

It is worth mentioning that the observed increase of particle
injection due to the precursor plasma heating is a consequence of the
thermal leakage model of particle injection adopted here. In this model,
a downstream particle, thermal or otherwise, with plasma frame speed
$v>u_2$, has a probability to return upstream which increases with $v$
\citep[see][for a discussion of the probability of returning
particles]{Bell78a}. That is, we assume that the subshock is transparent
to all particles with $v>u_2$, but only some of these particles get
injected, depending on their random histories. Particles that don't get
injected are convected far downstream out of the system. An alternative
model of injection \citep[see, for example,][]{BGV2005} is one where
only particles with a gyroradius greater than the shock thickness can
get injected. The assumption with this threshold injection model is that
the subshock thickness is comparable to the gyroradius of a downstream
thermal particle and only those particles with speeds $v > \xi
v_\mathrm{th}$ can be injected. Particles with $v < \xi v_\mathrm{th}$
are somehow blocked by the subshock.  Here, $v_\mathrm{th}$ is the
downstream thermal speed and $\xi$ is a free parameter, typically taken
to be between 2 and 4.

Despite being similar, it can be expected that these injection models
will react differently to precursor heating.
Namely, in the \citet{BGV2005} model the fraction of
injected particles may be insensitive to the precursor heating if $\xi$
is fixed\footnote{\citet{AB2006}, who performed a brief analysis of the
impact of the turbulence dissipation on the \NL\ shock structure in a
way similar to ours, did not report an influence of heating on the
injection rate.}, because the same particle injection rate occurs
regardless of the pre-subshock temperature $T_1$ and downstream
temperature $T_2$.
While both models are highly simplified
descriptions of the complex subshock
\citep[see, e.g.][]{Malkov98, GE2000}, they offer two scenarios for
grasping a qualitatively correct behavior of a shock where
particle injection and acceleration are coupled
to turbulence generation and flow modification.  Hopefully, a
clearer view of particle injection by self-generated turbulence in a
strongly magnetized subshock will become available when
relevant full particle PIC or hybrid
simulations are performed.

With the general trends observed here in mind,
we now show how \NL\ effects modify the effect dissipation has
on injection and MFA.

\subsection{Fully Nonlinear Model}

\label{nlresults}

In this section we demonstrate the results of the fully nonlinear
models, in which the flow structure, compression ratio, magnetic
turbulence, and particle distribution are all determined \SCly, so
that the fluxes of mass, momentum and energy are conserved across the
shock.

We use two sets of parameters, one with the far upstream gas
temperature $T_0=10^4$~K and the far upstream particle density $n_0 =
0.3$~\pcc, typical of the cold interstellar medium (ISM), and one with
$T_0=10^6$ K and $n_0=0.003$ \pcc, typical of the hot ISM.
In both cases we assumed the shock speed $u_0 = 5000$ \kmps, and the
initial magnetic field $B_0 = 3$ \muG\ (giving an equipartition of
magnetic and thermal energy far upstream, $n_0 k_B T_0 \approx
B_0^2/(8\pi)$). The corresponding sonic and \Alf\ Mach numbers are $M_s
\approx M_A \approx 400$ in both cases).  The size of the shocks
was limited by a FEB located at $\xFEB = -10^5 \rgzero\approx -3 \cdot
10^{-4}$ pc.
For both cases, we ran seven
simulations with different values of the dissipation rate $\heatpar$,
namely $\heatpar \in \{ 0; 0.1; 0.25; 0.5; 0.75; 0.9;
1.0\}$. Also, for the \hotism\ case we ran a
simulation neglecting the streaming instability effects, i.e., keeping
the magnetic field constant throughout the shock and assuming that the
precursor plasma is heated only by adiabatic compression (this model
will be referred to as the `no MFA case').

\clearpage
\begin{deluxetable}{ lrrrrrrr }
\tabletypesize{\small}
\tablecaption{Summary of Non-linear Simulation in a Cold ISM \label{sumnl_cold}}
\tablewidth{0pt}
\tablehead{
$\heatpar$         & 0.00  & 0.10  & 0.25  & 0.50  & 0.75  & 0.95  & 1.00
}
\startdata
$\rtot$            & 16.0  & 16.2  & 14.5  & 14.6  & 14.0  & 13.2  & 13.0  \\
$\rsub$            & 2.95  & 2.83  & 2.75  & 2.59  & 2.50  & 2.50  & 2.51  \\
$\Befftwo$, \muG   & 345   & 323   & 284   & 232   & 158   & 71    & 21    \\
$\Btrend$, \muG    & 345   & 327   & 299   & 245   & 174   & 80    & 21    \\
$\left<T(x<0)\right>$, $10^4$ K
                   & 1.06  & 4.3   & 9.0   & 17    & 26    & 37    & 56    \\
$T_1$, $10^4$ K    & 3.3   & 68    & 160   & 330   & 490   & 610   & 650   \\
$T_2$, $10^4$ K    & 1400  & 1500  & 1600  & 1600  & 1800  & 2000  & 2200  \\
$\Msone$           & 44    & 9.5   & 6.3   & 4.2   & 3.5   & 3.3   & 3.2   \\
$\fcr$, \%         & 1.0   & 1.2   & 1.6   & 2.1   & 2.6   & 3.1   & 3.2   \\
$\pmax / m_pc$     & 500   & 450   & 400   & 350   & 250   & 150   & 80    \\
$\left<\gamma(x<0)\right>$
                   & 1.33  & 1.33  & 1.33  & 1.33  & 1.34  & 1.34  & 1.34  \\
$\gamma_2$         & 1.38  & 1.38  & 1.38  & 1.39  & 1.39  & 1.40  & 1.41  \\
$\xtr  / \rgzero$  & -0.005&-0.001 & -0.001& -0.001&-0.002 & -0.004& -0.02 \\
$\xFP  / \rgzero$  & -0.04 &-0.06  & -0.06 & -0.09 & -0.23 &-0.57  & -2.1  \\
\enddata
\footnotesize\tablecomments{Here $\heatpar$, the fraction of dissipated energy, is the
  model input parameter, and the rest are results of the self-consistent
  simulation, as follows: $\rtot=u_0/u_2$ is the total shock compression ratio,
  $\rsub=u_1/u_2$ is the subshock compression, $\Befftwo$ is the amplified
  effective magnetic field downstream, $\Btrend$ is the trend value
  calculated from Eq.~(\ref{btrend}), $\left<T(x<0)\right>$ is the temperature
  of the precursor averaged over the volume from $x=\xFEB$ to $x=0$,
  $T_1$ is the temperature at $x=\xtr$, $T_2$ is the volume-averaged
  temperature at $x>0$ (all temperatures are calculated from the ideal
  gas law~(\ref{idealgaslaw})), $\Msone$ is the sonic Mach number at $x=\xtr$,
  $\fcr$ is the fraction of injected thermal particles, $\pmax$ is the
  maximum particle momentum (i.e., the momentum at which $f(p)$ starts falling
  off exponentially with $p$), $\left<\gamma(x<0)\right>$ is the value of
  the polytropic index of  particle gas, calculated from particle pressure
  and internal energy density as described in Appendix~\ref{compressionratio},
  and averaged over volume from $x=\xFEB$ to $x=0$, $\gamma_2$ is the
  same quantity averaged over $x>0$, $\xtr$ is the point at which the
  subshock starts, as defined by Eq.~(\ref{subshocklocation}),
  and $\xFP$ is the point defined by Eq.~(\ref{xfastpush}), at which
  thermal particles were introduced by the APA procedure as described
  in Appendix~\ref{fastpush}.
}
\end{deluxetable}

%\clearpage

\begin{deluxetable}{ lrrrrrrrr }
\tabletypesize{\small}
\tablecaption{Summary of Non-linear Simulation in a Hot ISM \label{sumnl_hot}}
\tablewidth{0pt}
\tablehead{
$\heatpar$         & 0.00  & 0.10  & 0.25  & 0.50  & 0.75  & 0.95  & 1.00  & No MFA
}
\startdata
$\rtot$            & 8.1   & 8.2   & 8.3   & 8.0   & 7.8   & 7.4   & 7.3   & 13     \\
$\rsub$            & 2.92  & 2.75  & 2.55  & 2.44  & 2.22  & 2.15  & 2.12  & 2.75   \\
$\Befftwo$, \muG   & 62    & 60    & 55    & 44    & 32    & 17    & 14    & 21     \\
$\Btrend$, \muG    & 62    & 59    & 54    & 45    & 33    & 19    & 13    & -      \\
$\left<T(x<0)\right>$, $10^6$ K
                   & 1.04  & 1.3   & 1.7   & 2.3   & 3.1   & 3.9   & 4.2   & 1.1    \\
$T_1$, $10^6$ K    & 2.0   & 6.0   & 13    & 23    & 34    & 42    & 43    & 2.7    \\
$T_2$, $10^6$ K    & 53    & 49    & 47    & 55    & 62    & 72    & 75    & 22     \\
$\Msone$           & 10.9  & 5.8   & 3.7   & 2.6   & 2.1   & 1.9   & 1.9   & 4.7    \\
$\fcr$, \%         & 1.2   & 1.6   & 2.5   & 4.0   & 6.4   & 6.9   & 6.4   & 2.4    \\
$\pmax / m_pc$     & 150   & 120   & 110   & 100   & 90    & 70    & 60    & 80     \\
$\left<\gamma(x<0)\right>$
                   & 1.34  & 1.34  & 1.34  & 1.34  & 1.34  & 1.34  & 1.35  & 1.34   \\
$\gamma_2$         & 1.43  & 1.43  & 1.43  & 1.44  & 1.45  & 1.45  & 1.45  & 1.41   \\
$\xtr  / \rgzero$  & -0.04 &-0.02  & -0.02 & -0.02 &-0.03  & -0.07 & -0.05 & -0.02  \\
$\xFP  / \rgzero$  & -0.1  &-0.1   & -0.2  & -0.2  & -0.4  & -0.9  & -1.4  & -0.1   \\
\enddata
\footnotesize\tablecomments{See the note for Table~\ref{sumnl_cold} for
the explanation of listed quantities.}
\end{deluxetable}
\clearpage

Referring to Tables~\ref{sumnl_cold} and \ref{sumnl_hot}, we summarize
some of the results of these models.
The effect of the turbulence dissipation into the thermal plasma is
evident in the values of the pre-subshock temperature $T_1$, the
downstream temperature $T_2$, and the volume-averaged precursor
temperature $\left< T(x<0) \right>$ (the averaging takes place between
$x=\xFEB$ and $x=0$).
The temperatures were calculated from the thermal particle pressure
$\Pth(x)$ using the ideal gas law equation~(\ref{idealgaslaw}).
The value of $T_1$ depends drastically on the
level of the turbulence dissipation $\heatpar$, increasing from
$\heatpar=0$ to $\heatpar=0.5$ by a factor of 100 in the
\coldismNoT case, and by a factor of 11 in the
\hotismNoT case (less in the latter case, because
the efficiency of the CR streaming instability in generating the
magnetic turbulence is less for the smaller $M_s$ and $M_A$ for
$T=10^6$~K).
The values of the temperature as high as $T_1$ are achieved upstream
only near the subshock; the volume-averaged upstream temperature,
$\left< T(x<0) \right>$, is significantly lower. The factors by which
$\left< T(x<0) \right>$ increases in the above
cases are 17 and 2.3, respectively.
The value of $\Mszero^2 / \Mazero$ is large in our models,
and the estimate~(\ref{upsilonequation}) predicts that
even a small amount of dissipation is enough
to raise the precursor temperature significantly. This is confirmed
by our results: even $\heatpar=0.1$ is enough to raise
the pre-subshock temperature $T_1$ approximately 20 times in the
\coldism\ case.

The downstream temperature, $T_2$, varies less with changing $\heatpar$,
because it is largely determined by the compression at the subshock,
which is controlled by many factors as we discuss below.
It is worth mentioning the case without
MFA reported in Table~\ref{sumnl_hot}.  Besides having a much larger
compression factor than the shocks with MFA ($\rtot=13$ as opposed to
$\rtot\lesssim 8$), it has a much smaller downstream temperature
($T_2=2.2\cdot 10^7$~K as opposed to $T_2 \gtrsim 5.3\cdot 10^7$~K)
These effects of dissipation on the precursor temperature may be
observable.

%\clearpage
\begin{figure}
\epsscale{1.0}
\plotone{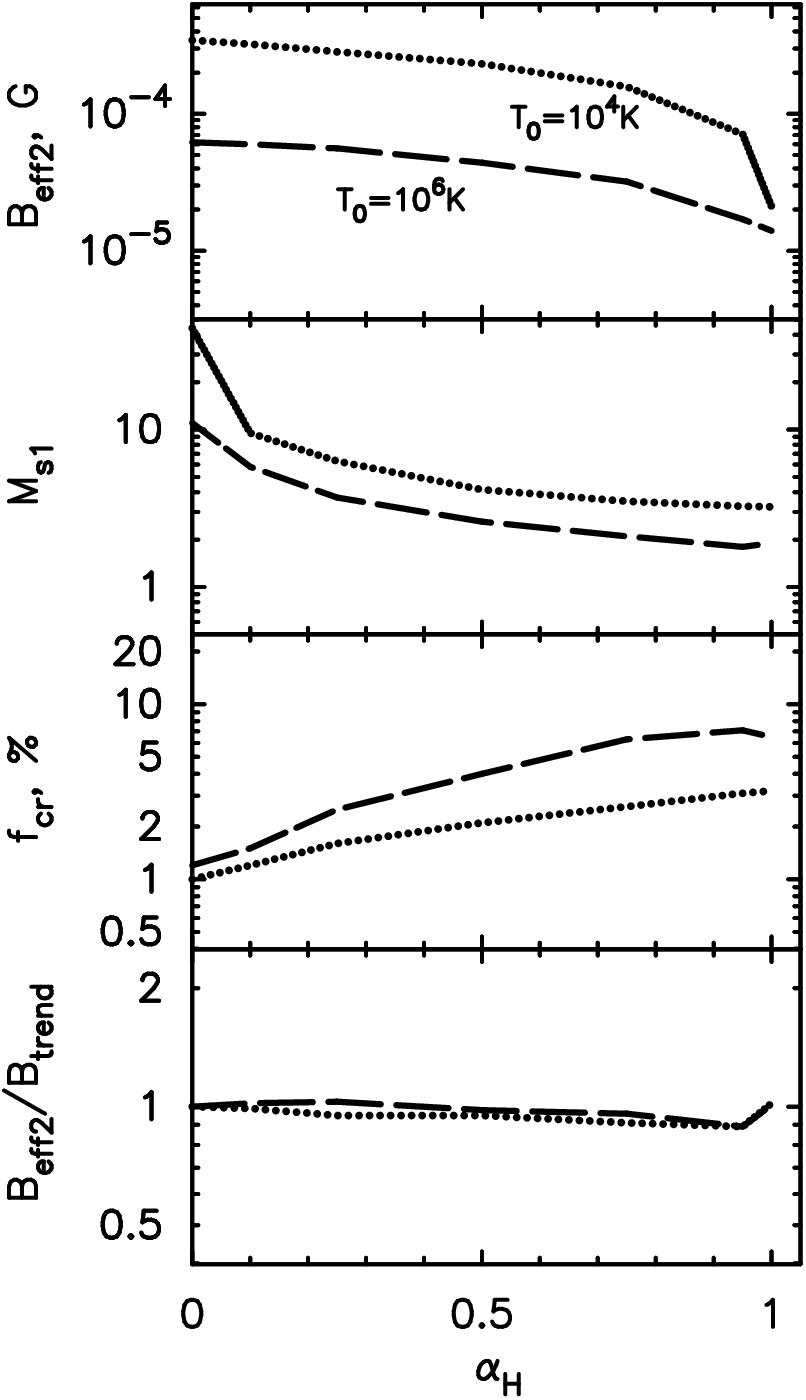}
\caption{Dependence of particle injection and magnetic field
  amplification on the rate of magnetic turbulence dissipation in the
  non-linearly modified shock with $u=5000$ \kmps\ in the \coldismNoT\
  and the \hotismNoT\ cases (the fully self-consistent models).  The
  $x$-axis variable is the turbulence dissipation rate $\heatpar$
  (constant throughout a shock), and the plotted quantities are the
  amplified downstream effective magnetic field $\Befftwo$, the subshock
  sonic Mach number $\Msone$, the fraction of
  simulation particles injected into the acceleration process
  $\fcr$, and the ratio of amplified field to $\Btrend$ downstream
  from the shock.
  \label{fignlsum}}
\end{figure}
%\clearpage

In Figure~\ref{fignlsum} we show results for $\fcr$, $\Msone$,
$\Befftwo$, and $\Btrend$ which can be compared to the results for
unmodified shocks shown in Figure~\ref{fig_unmod_inj}. For the modified
shocks, the fraction of the thermal particles crossing the shock
backwards for the first time, $\fcr$, clearly increases by a large
factor with $\heatpar$, which can be explained by the value of $\Msone$
dropping quickly below 10. One could expect that the amplified
effective magnetic field $\Befftwo$ would behave similarly to the
$\rtot=3.5$ case in Section~\ref{injectionvsdissipation}, i.e. that
$\Befftwo$ would not decrease or even would increase for larger
$\heatpar$.  Instead, $\Befftwo$ behaves approximately according to the
trend~(\ref{btwotrend}), as the values of $\Btrend$ from
Tables~\ref{sumnl_cold} and \ref{sumnl_hot} show and the bottom panel of
Fig.~\ref{fignlsum} illustrates. The important point is that, even
though precursor heating causes the {\it injection efficiency} to
increase substantially, the {\it efficiency of particle acceleration}
and magnetic turbulence generation is hardly changed.
We base this assertion on the fact that $\Befftwo$ remains
close to $\Btrend$, which was derived under the assumption that changing
$\heatpar$ preserves the total energy generated by the instability, but
re-distributes it between the turbulence and the thermal particles. The
fact that the particle acceleration efficiency is insensitive to
$\heatpar$ is directly seen in the results displayed in
Figure~\ref{fig_fnp_1e4_1e6} below.

Considering how much the injection rate $\fcr$ increases with
$\heatpar$, and how much the upstream temperature
of the thermal plasma, $T_1$, is affected by the heating,
it is somewhat surprising that the
trend of the amplified effective field $\Befftwo$ is unaffected. The
mechanism by which the shock adjusts to the changing heating and
injection in order to preserve the MFA efficiency can be understood by
looking at the trend of the total compression ratio $\rtot$ and the
subshock compression ratio $\rsub$ in Tables~\ref{sumnl_cold} and
\ref{sumnl_hot}: they both decrease significantly for higher
$\heatpar$.
The decrease in $\rsub$ is easy to understand:
with the turbulence dissipation operating in the precursor
$\Msone$ goes down, which lowers $\rsub$. Additionally,
decreasing $\Pwone$ helps to reduce $\rsub$, and
with a boost
of the particle injection rate, the particles returning for the first
time increase in number and build up some extra pressure just upstream
of the shock, which causes
the flow to slow down in that region, thus reducing the ratio
$\rsub$.
The decrease in $\rtot$ results from more complex processes. Here,
the histories of `adolescent' particles, i.e., those particles
returning upstream for the first time or for the first few times, are
critical. Adolescent particles, while superthermal, are still highly
anisotropic in the shock frame and how they get accelerated in the
smoothed precursor just upstream of the subshock determines the number
and energies of the `mature' superthermal particles, i.e., those
particles with enough energy to be nearly isotropic in the shock
frame. The mature particles determine the CR pressure and precursor
smoothing on larger scales.

\begin{figure}
\epsscale{1.00} \plottwo{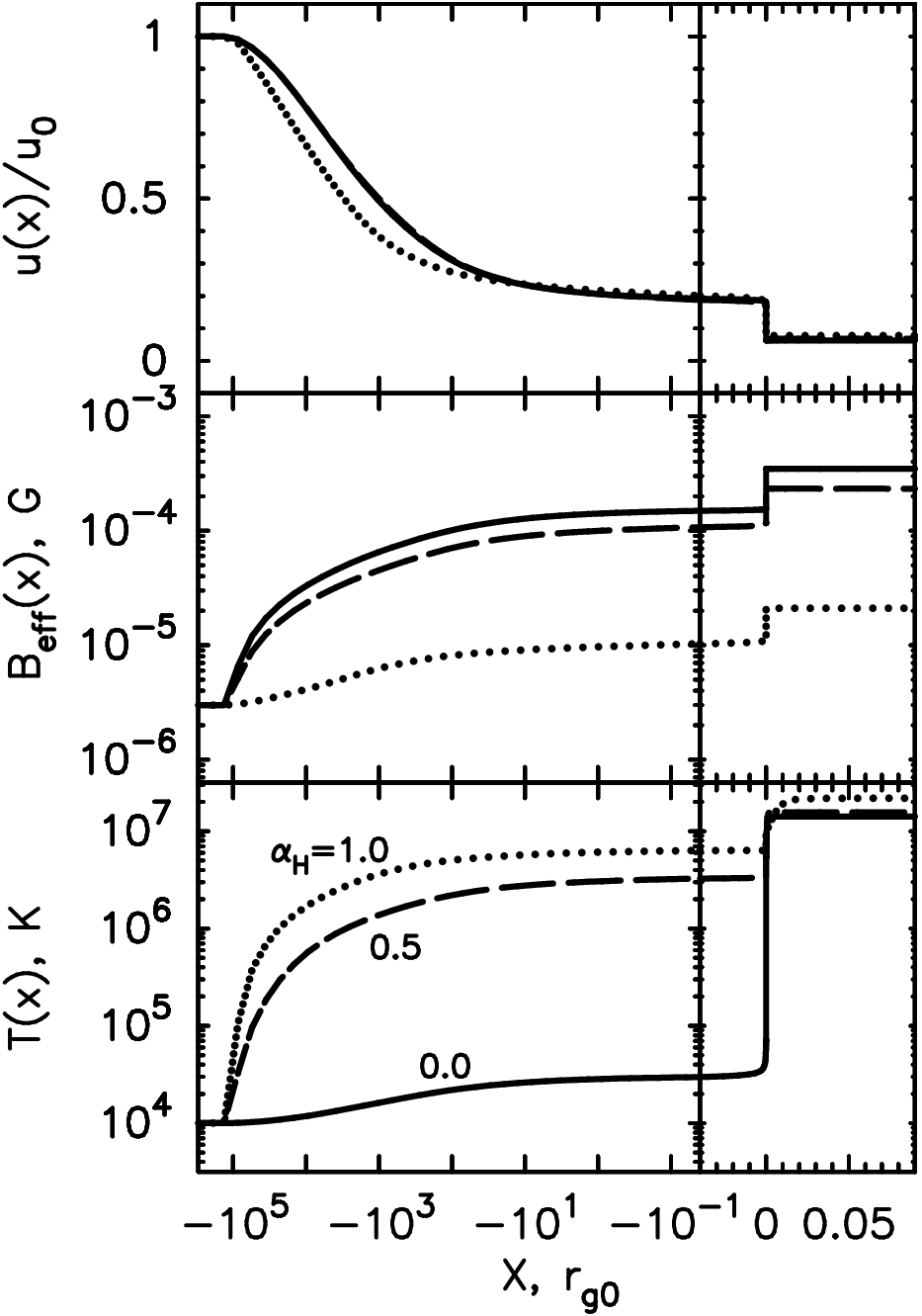}{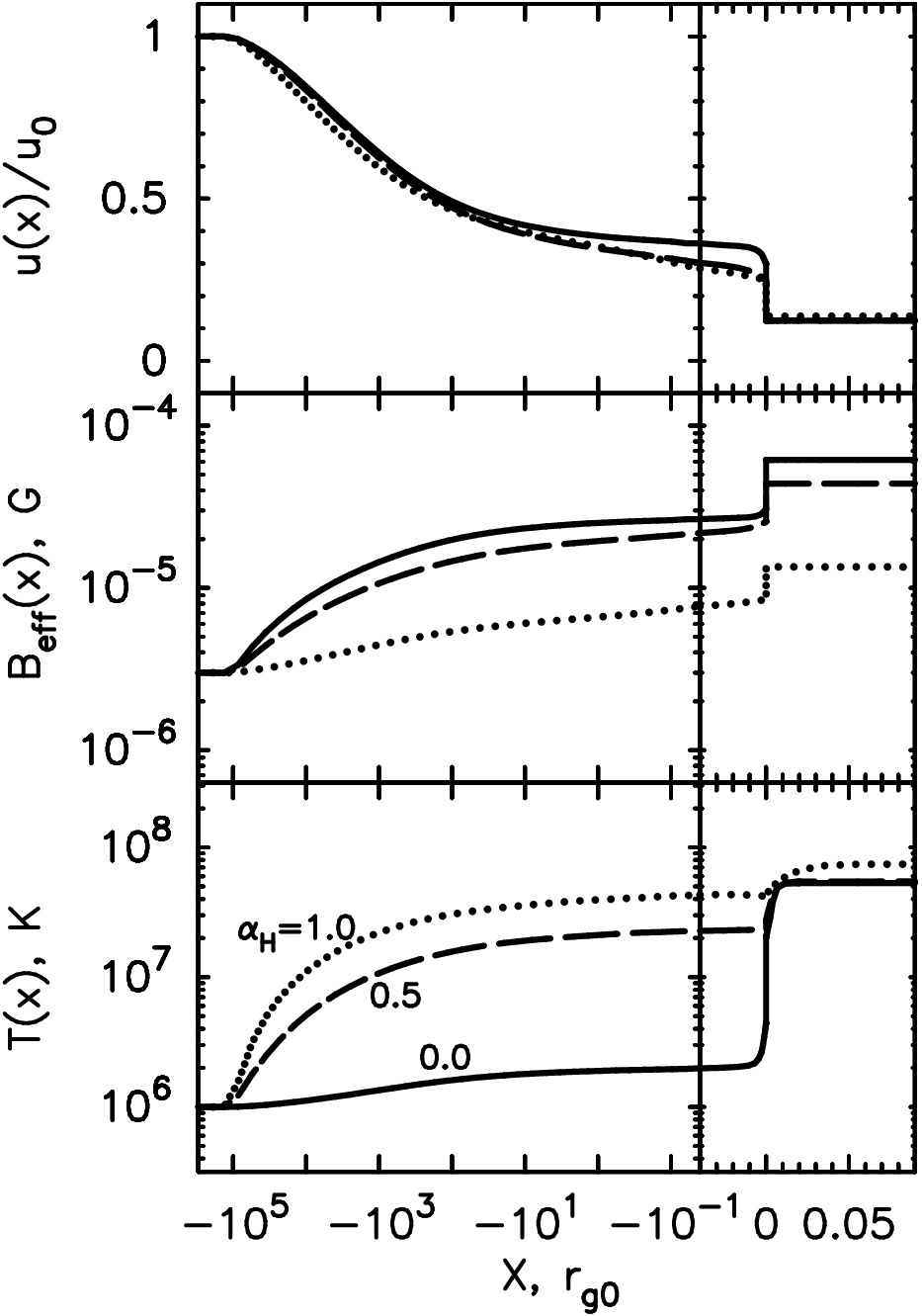}
\caption{ Results of non-linear simulations in the \coldismNoT\
(left) and \hotismNoT\ (right) with different values of
$\heatpar$.  The solid, dashed and dotted lines correspond,
respectively, to $\heatpar = 0$, $0.5$ and $1.0$.  The plotted
quantities are the bulk flow speed $u(x)$, the effective amplified
magnetic field $\Beff(x)$ and the thermal gas temperature $T(x)$. The
shock is located at $x=0$, and note the change from the logarithmic to
the linear scale at $x=-0.05\:\rgzero$.
  \label{fig_ubt_1e4_1e6}}
\end{figure}

\begin{figure}
\epsscale{1.00} \plotone{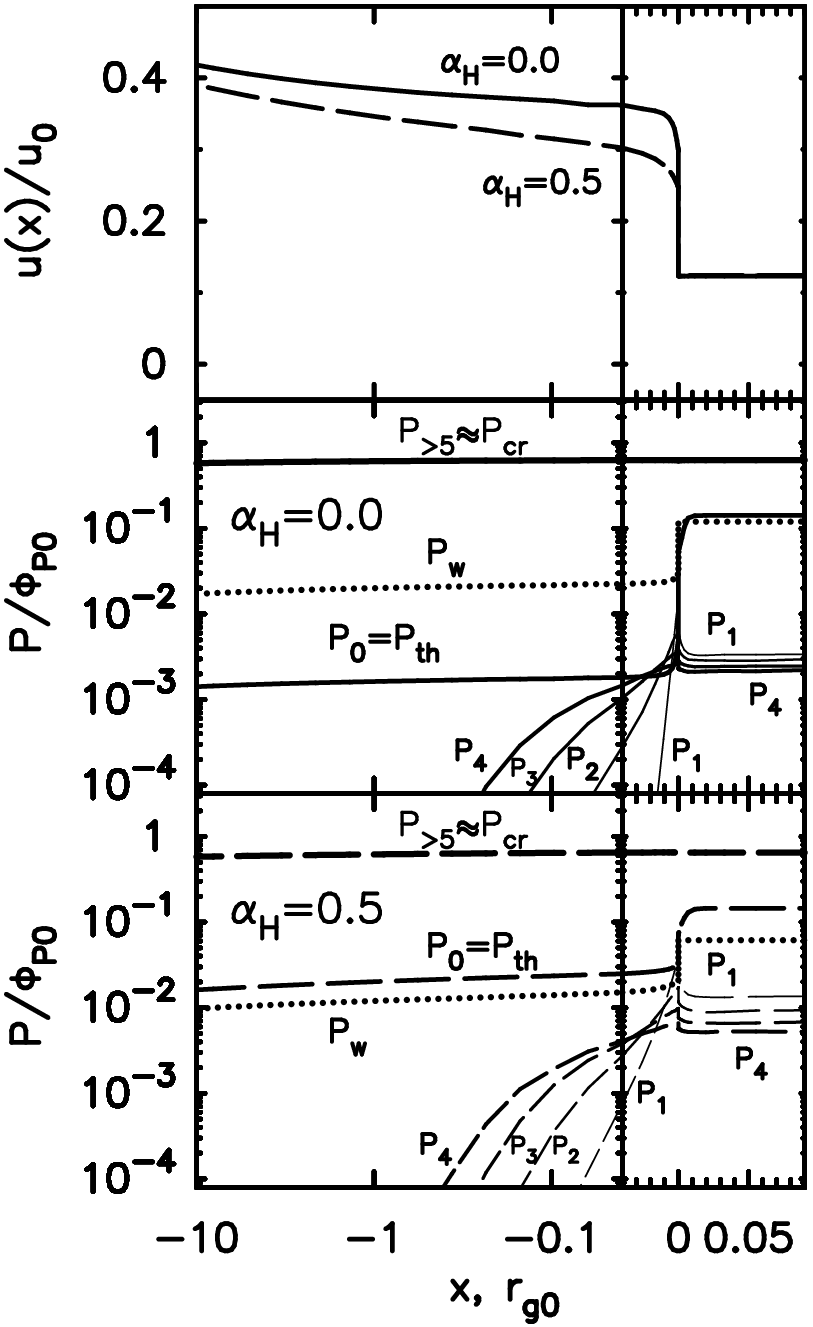}
\caption{Enlarged subshock region in the \hotismNoT\ case.  The top
  panel shows the flow speed normalized to $u_0$ for the
  $\heatpar=0.0$ and $\heatpar=0.5$ models. The middle panel shows
  various constituents of pressure for the $\heatpar=0.0$ run: the
  thermal pressure $\Pth$, the pressures of `adolescent' particles that
  crossed the shock 1, 2, 3 and 4 times ($P_1$, $P_2$, $P_3$ and $P_4$,
  respectively), the pressure of particles that have crossed 5 and more
  times $P_\mathrm{>5}$ and the magnetic turbulence pressure $P_w$, all
  of the above are normalized to the far upstream momentum flux
  $\Phi_\mathrm{P0}$. The bottom panel shows the same quantities for
  $\heatpar=0.5$.
\label{fig_partpress}}
\end{figure}

Further understanding of the shock adjustment to the changing
dissipation can be gained by studying Figures \ref{fig_ubt_1e4_1e6} -
\ref{fig_fnp_1e4_1e6}, in which we plot the spatial
structure and the momentum-dependent quantities of the shocks in the
\coldismNoT\ and the \hotismNoT\ cases for $\heatpar \in \{0; 0.5; 1\}$.

%\clearpage

%\clearpage

Figure~\ref{fig_ubt_1e4_1e6} shows an overlap in the curves for the flow
speed $u(x)$ in the $\heatpar=0$ and $\heatpar=0.5$ models, and only
close to the subshock $u(x)$ falls off more rapidly towards the subshock
in the $\heatpar=0.5$ case, resulting eventually in a lower
$\rsub$. This means that for the high energy particles, which diffuse
far upstream, the acceleration process will go on in about the same
way with and without moderate turbulence dissipation
(and the acceleration efficiency will be preserved with changing
$\heatpar$).
For lower energy particles, however, there will be observable differences in the
energy spectrum.  The $\heatpar=1.0$ case has a significantly smoother
precursor, which is not unusual, given the lower maximal energy of the
accelerated particles in this case (because of the magnetic field
remaining low).  The thermal gas temperatures $T(x)$ plotted in the
bottom panels of Figure~\ref{fig_ubt_1e4_1e6} were calculated from the
thermal pressure $\Pth(x)$ using~(\ref{idealgaslaw}) and show that the
temperature becomes high well in front of the subshock.

In Figure~\ref{fig_partpress} the subshock region for the \hotismNoT\
case is shown enlarged, and we can compare details of the models with
($\heatpar=0.5$) and without dissipation ($\heatpar=0.0$).  In the
absence of turbulence dissipation, the thermal pressure $\Pth$ remains
low upstream (see the middle panel), and the subshock transition is
dominated by the magnetic pressure $P_w$.
For $\heatpar=0.5$
(the bottom panel) the thermal pressure $\Pth$ just before the shock
increases enough to become comparable with the magnetic pressure, but
also the heating-boosted particle injection brings up the pressures of
the `adolescent' particles. As the plot shows, for $\heatpar=0.5$ the
pressures produced by the first and second time returning particles
($P_1$ and $P_2$) are not small compared to $\Pth$ and $P_w$ just
upstream of the shock, which contributes to the reduction of $\rsub$
described above. However, the pressure of the `mature' particles,
$P_\mathrm{>5}$, doesn't change much, which is a result of the
non-linear response of the shock structure to the increased injection.

%\clearpage
\begin{figure}
\epsscale{1.00}
\plottwo{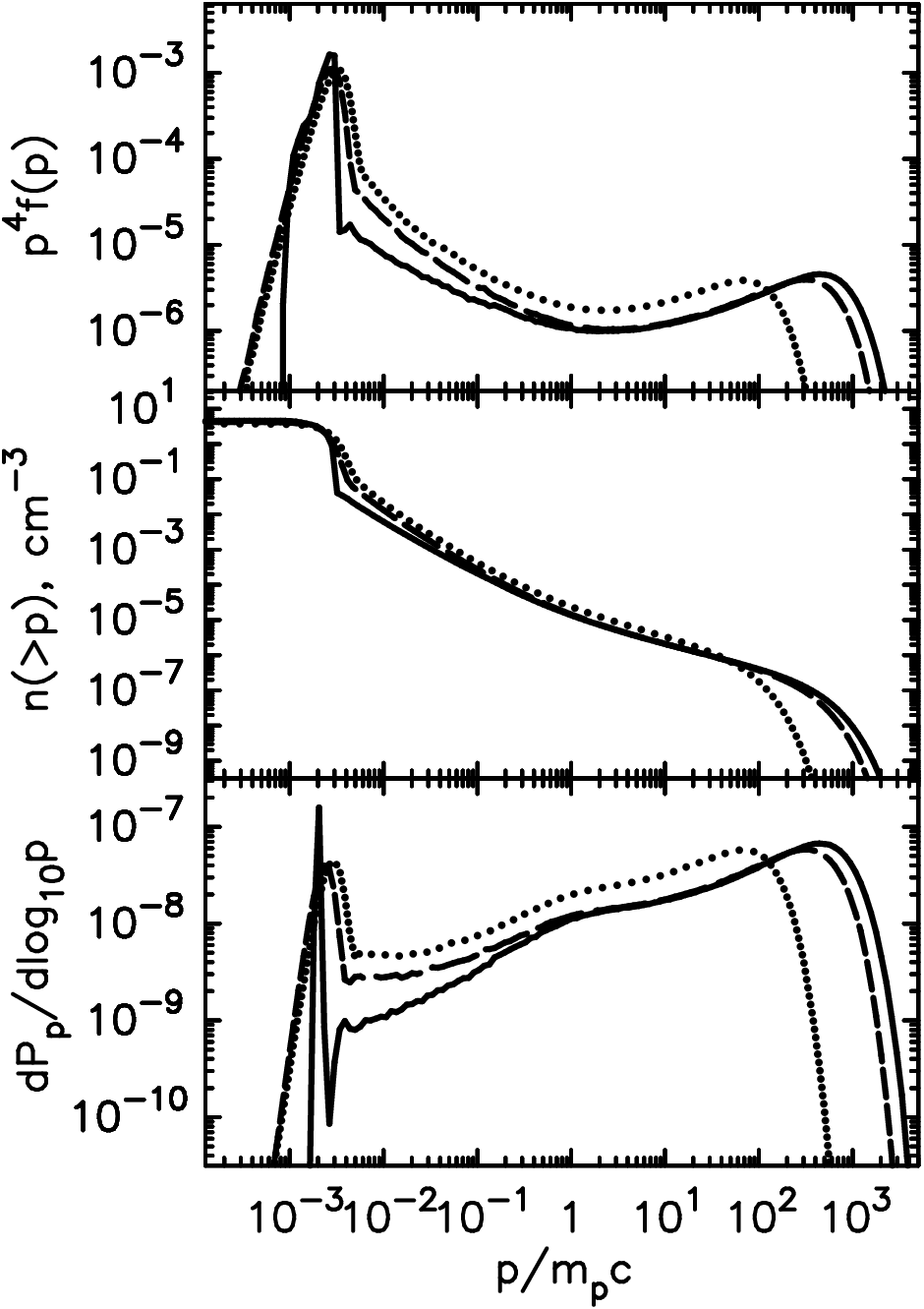}{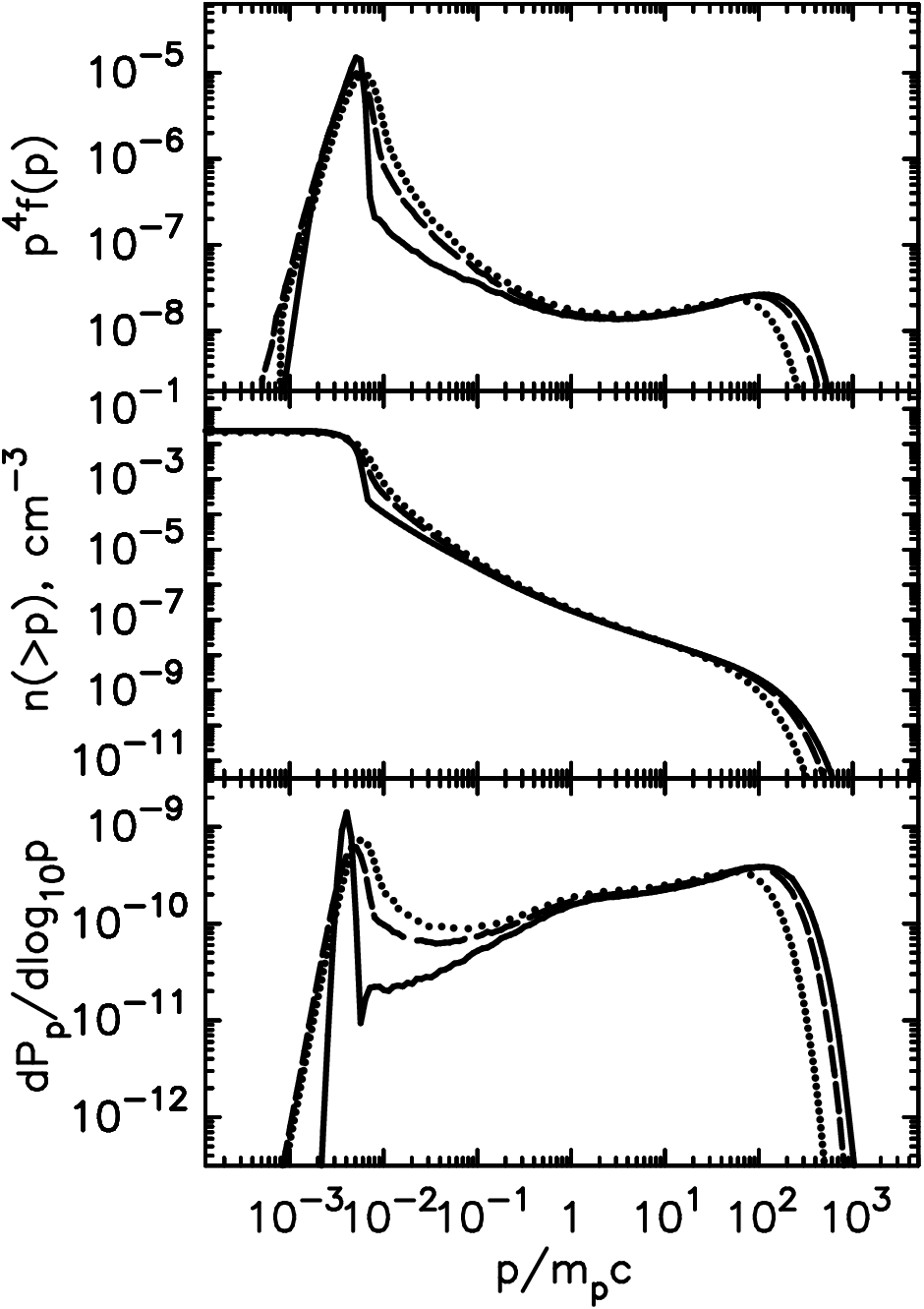}
\caption{Results of non-linear simulations in the \coldismNoT
  (left) and \hotismNoT\ (right) with different values
  of $\heatpar$.  Line styles as in Fig.~\ref{fig_ubt_1e4_1e6}. The
  plotted quantities are: the particle distribution function in the
  shock frame $f(p)$ multiplied by $p^4$ (the normalization is
  such that $\int{4\pi p'^2 f(p') dp'}=n$, $n$ being the number density
  in \pcc, and $p'=p/(m_p c)$), the number of particles $n(>p)$ with
  momentum greater than $p$ (in \pcc) and the particle pressure (in
  dynes per cm$^2$) per decade of normalized momentum
  $dP_p/d\log_\mathrm{10}{p/(m_p c)}$. All quantities are calculated
  downstream at $x=+6 \rgzero$.
     \label{fig_fnp_1e4_1e6}}
\end{figure}
%\clearpage

The low energy parts of the particle distribution functions shown in
Figure~\ref{fig_fnp_1e4_1e6} are significantly different for models with
and without dissipation in both the \coldismNoT\ and the \hotismNoT\
cases. The apparent widening of the thermal peak reflects the increase
in the downstream gas temperature $T_2$. The differences extend from the
thermal peak to mildly superthermal momenta $0.2\: m_p c$, which shows
an increased population of the `adolescent' particles with speeds up to
$v\approx 0.2 c \approx 12 u_0$ when the turbulence dissipation
operates. The high energy ($p>0.2 \: m_p c$) parts of the spectra for
$\heatpar=0$ and $\heatpar=0.5$ are similar (except
for a lower $\pmax$ due to a lower value of the amplified field in the
$\heatpar=0.5$ case), confirming our
assertion about the preservation of the particle acceleration
efficiency. The increased population of the low-energy particles just
above the thermal peak should influence the shock's X-ray emission.

The characteristic concave curvature of the particle spectra above
the thermal peak is clearly seen in the top panels of
Figure~\ref{fig_fnp_1e4_1e6}. These shocks are strongly \NL\ and, as
the bottom panels in Figure~\ref{fig_fnp_1e4_1e6} show, most of the
pressure is in the highest energy particles. For these examples, 60 to
80 percent of the downstream momentum flux is in CR particles. The
number of particles producing this pressure is small, however, and as
the plots in the middle panels show, the fraction of particles above
the thermal peak is on the order of $10^{-3}$, and the fraction of
particles above 1~GeV is around $10^{-6}$ in all cases. In addition to
the pressure (and energy) in the distributions shown, a sizable
fraction of shock ram kinetic energy flux escapes at the FEB.

To summarize,
for both the unmodified (Fig.~\ref{fig_unmod_inj}) and modified
(Fig.~\ref{fignlsum}) cases, $\Msone$ drops and
$\fcr$ grows as $\heatpar$ increases. The surprising result is that
$\Befftwo$ can increase in the unmodified shock as $\heatpar$ goes up if
$\rtot$ is large enough. This indicates that the boosted injection
efficiency (i.e., larger $\fcr$) outweighs the effects of field
damping. This doesn't happen in the modified case (top panel of
Fig.~\ref{fignlsum}) because of the \NL\ effects from the increased
injection. From Fig.~\ref{fig_partpress} we see that the boosted
injection results in a smoother subshock and this makes it harder for
low energy adolescent particles to gain energy. Once particles reach a
high enough momentum ($p\gtrsim 0.2 m_pc$; see the top panel of
Fig.~\ref{fig_fnp_1e4_1e6}) they are able to diffuse far enough upstream
where the boost in injection has a lesser effect.

We must emphasise again that these results are very sensitive to
the physics of particle injection at the subshock. It is difficult to
predict how the \NL\ results would change if a different model of
injection was used, but we can refer the reader to the work of
\citet{AB2006}, who performed a similar calculation using the
threshold injection model with a different diffusion coefficient.

The free escape boundary in our simulation was relatively close to the
shock ($\xFEB = -10^5 \rgzero \approx - 3 \times 10^{-4}$ pc), and the
maximum accelerated particle energy was on the order of hundreds of
GeV. These quantities, chosen to save computation time, are
several orders of magnitude short of typical SNR values.
Nevertheless, we don't believe our results will be changed
qualitatively if $\pmax$ is increased. The reason is that even with our
relatively low $\pmax$, the fraction of internal energy in \rel\ particles is
still large and the volume-averaged value of the
polytropic index of the precursor plasma, $\left<\gamma(x<0)\right>$,
as shown in Tables~\ref{sumnl_cold} and \ref{sumnl_hot}, is much closer to
the $4/3$ of a fully relativistic gas than to $5/3$ for a
nonrelativistic one. Increasing $\pmax$ will not lower the
polytropic index of the gas any further and, consequently, the plasma
compression and the subsequent acceleration efficiency will not
change significantly \citep[see][]{BE99}.

\section{Conclusions}

\label{conclusions}

We have parameterized magnetic turbulence dissipation as a fraction of
turbulence energy generation and included this effect in our \MC\ model
of strongly \NL\ shocks undergoing efficient DSA. The energy removed
from the turbulence goes directly into the thermal particle population
in the shock precursor. The \mc\ simulation \SCly\ reacts to the changes
in precursor heating and adjusts the injection of thermal particles into
the DSA mechanism, as well as other \NL\ effects of DSA, accordingly.

Our two most important results are, first, that even a small rate
($\sim 10$\%) of turbulence dissipation can drastically increase the
precursor temperature, and second, that the precursor heating boosts
particle injection into DSA by a large factor. The increase in particle
injection modifies the low-energy part of the particle spectrum but, due
to \NL\ feedback effects, does not significantly change the overall
efficiency or the high energy part of the spectrum. Both the precursor
heating and modified spectral shape that occur with dissipation may have
observable consequences.

The fact that the shock back-reaction to the increased injection
prevents the acceleration efficiency from
changing significantly is a clear consequence of the non-linear
structure of the system. The boosted particle injection additionally
smoothes the flow speed close to the subshock, which makes it harder for
particles returning upstream to gain energy. As a result, the population
of the high energy particles is not much changed
(except for the decrease in the maximum particle momentum due to the
reduction in the effective amplified magnetic field from dissipation)
and, because those particles carry the bulk of the CR
pressure $\Pcr$ which drives the streaming
instability, the amplification of the magnetic field is not
strongly affected by the heating-boosted particle injection.

The parameterization we use here is a simple one and a more advanced
description of the turbulence damping may change our results. In our
model the energy drained from the magnetic turbulence, at all
wavelengths, is directly `pumped' into the thermal
particles. Superthermal particles only gain extra
energy due to heating because the thermal particles were more likely to
return upstream and get accelerated.  In a more advanced model of
dissipation, where energy cascades from large-scale turbulence harmonics
to the short-scale ones, the low energy CRs might gain energy directly
from the dissipation. It is conceivable that cascading effects might
increase the overall acceleration efficiency, the magnetic field
amplification, and increase the maximum particle energy a shock can
produce.

It is also possible that non-resonant
turbulence instabilities play an important role in magnetic
field amplification \citep[e.g.,][]{PLM2006}.  This opens another
possibility for the turbulence dissipation to produce an increase in the
magnetic field amplification.  For instance, \citet{BT2005} proposed a
mechanism for generating long-wavelength
perturbations of magnetic fields by low energy particles. If such a
mechanism is responsible for generation of a significant fraction of the
turbulence that confines the highest energy particles, then the
increased particle injection due to the precursor heating may raise the
maximum particle energy and, possibly, the value of the amplified
magnetic field.

While our model is for the most part phenomenological as far
as particle transport, injection and acceleration, magnetic field
generation, and dissipation are concerned, it allows us to investigate
the coupled nonlinear effects in a shock undergoing efficient DSA and
MFA. The more efficient DSA is, the more basic considerations of
momentum and energy conservation determine the shock structure and our
model describes these effects fully.

\acknowledgments

We thank the anonymous referee for a number of very valuable comments and suggestions.
D.~C.~E. and A.~V. wish to acknowledge support from NASA grants NNH04Zss001N-LTSA
and 06-ATP06-21. A.~M.~B. acknowledges support from RBRF grant 06-02-16884. 

% NASA ATP ( NNX07AG79G )

\appendix

\section{A. Comments on PIC simulations of MFA}

\label{estimatesforpic}

There are two basic reasons why the problem of MFA in nonlinear
diffusive shock acceleration (NL-DSA) is particularly difficult for
particle-in-cell (PIC) simulations. The
first is that PIC simulations must be done fully in three dimensions to
properly account for cross-field diffusion. As \citet{JJB98} proved from
first principles, PIC simulations with one or more ignorable dimensions
unphysically prevent particles from crossing magnetic field lines.  In
all but strictly parallel shock geometry,\footnote{Parallel geometry is
where the upstream magnetic field is parallel to the shock normal.} a
condition which never occurs in strong turbulence, cross-field
scattering is expected to contribute importantly to particle injection
and must be fully accounted for if injection from the thermal background
is to be modeled accurately.

The second reason is that, in \nonrel\ shocks, NL-DSA spans large
spatial, temporal, and momentum scales. The range of scales is more
important than might be expected because DSA is intrinsically efficient
and \NL\ effects tend to place a large fraction of the particle pressure
in the highest energy particles (see
Fig.~\ref{fig_fnp_1e4_1e6}). The highest energy particles, with the
largest diffusion lengths and longest acceleration times, feedback on
the injection of the lowest energy particles with the shortest scales.
The accelerated particles exchange their momentum and energy with the
incoming thermal plasma through the magnetic fluctuations coupled to the
flow. This results in the flow being decelerated and the plasma being
heated. The structure of the shock, including the subshock where fresh
particles are injected, depends critically on the highest energy
particles in the system.

A plasma simulation must resolve the electron skin depth,
$c/\ElPlasmafreq$, i.e., $\Lcell < c/\ElPlasmafreq$, where
$\ElPlasmafreq=[4 \pi n_e e^2/m_e]^{1/2}$ is the electron plasma
frequency and $\Lcell$ is the simulation cell size. Here, $n_e$ is
the electron number density, $m_e$ is the electron mass and $c$ and
$e$ have their usual meanings. The simulation must
also have a time step small compared to $\ElPlasmafreq^{-1}$, i.e.,
$\tstep < \ElPlasmafreq^{-1}$.
If we wish to follow the acceleration of protons in DSA to the TeV
energies present in SNRs we must have a simulation box that is as large
as the upstream diffusion length of the highest energy protons, i.e.,
$\kappa(\mathrm{\Emax})/u_0 \sim r_g(\Emax) c / (3 u_0)$, where $\kappa$
is the diffusion coefficient, $r_g(\Emax)$ is the gyroradius of a \rel\
proton with the energy $\Emax$, $u_0$ is the shock speed, and we have
assumed Bohm diffusion.
The simulation must also be able to run for as long as the acceleration
time of the highest energy protons, $\tacc(\Emax) \sim \Emax c /(eBu_0^2)$.
Here, $B$ is some average magnetic field.
The spatial condition gives
\begin{equation}
\frac{\kappa(E_\mathrm{max})/u_0}{(c/\ElPlasmafreq)} \sim 6\xx{11}
\left ( \frac{\Emax}{\mathrm{TeV}} \right )
\left ( \frac{u_0}{1000 \, \mathrm{km \, s}^{-1}} \right )^{-1}
\left ( \frac{B}{\mu \mathrm{G}} \right )^{-1}
\left ( \frac{n_e}{\mathrm{cm}^{-3}}\right )^{1/2}
\left ( \frac{f}{1836}\right )^{1/2}
\ ,
\end{equation}
for the number of cells {\it in one dimension}. The factor $f=m_p/m_e$
is the proton to electron  mass ratio. From the acceleration
time condition, the required number of time steps is,
\begin{equation}
\frac{\tacc(E_\mathrm{max})}{\ElPlasmafreq^{-1}} \sim 6\xx{14}
\left ( \frac{\Emax}{\mathrm{TeV}} \right )
\left ( \frac{u_0}{1000\,\mathrm{km\,s}^{-1}}\right )^{-2}
\left ( \frac{B}{\mu \mathrm{G}} \right )^{-1}
\left ( \frac{n_e}{\mathrm{cm}^{-3}}\right )^{1/2}
\left ( \frac{f}{1836}\right )^{1/2}
\ .
\end{equation}
Even with $f=1$ these numbers are obviously far beyond any conceivable
computing capabilities and they show that approximate methods are
essential for studying NL-DSA.

One approximation that is often used is a hybrid PIC simulation where
the electrons are treated as a background fluid. To get the 
estimate of the requirements in this case we can take the
minimum cell size as the thermal proton gyroradius, 
$\rgzero = c \sqrt{2 m_p \Eth}/(e B)$.
Now, the number of cells, {\it again in one dimension,} is: 
\begin{equation}
\label{eq:cell}
\frac{\kappa(\Emax)/u_0}{\rgzero} \sim 7\xx{7}
\left ( \frac{\Emax}{\mathrm{TeV}} \right )
\left ( \frac{u_0}{1000 \, \mathrm{km \, s}^{-1}} \right)^{-1}
\left ( \frac{\Eth}{\mathrm{keV}} \right )^{-1/2}
\ .
\end{equation}
The time step size must be $\tstep < \protongyrofreq^{-1}$, 
where $\protongyrofreq=eB/m_{p}c$ is the thermal proton gyrofrequency.
This gives the number of time steps to reach 1 TeV,
\begin{equation}
\label{eq:step}
\frac{\tacc(\Emax)}{\protongyrofreq^{-1}} \sim 1\xx{8}
\left ( \frac{\Emax}{\mathrm{TeV}} \right )
\left ( \frac{u_0}{1000\,\mathrm{km\,s}^{-1}}\right )^{-2}
\ .
\end{equation}
These combined spatial and temporal requirements, even for the most
optimistic case of a hybrid simulation with an unrealistically large
$\tstep$, are well beyond existing computing capabilities unless
%exceptionally clever shortcuts are devised or 
a maximum energy well below 1 TeV is used.

Since the three-dimensional requirement is fundamental and relaxing it
eliminates cross-field diffusion, restricting the energy range is the
best way to make the problem assessable to hybrid PIC
simulations. However, since producing \rel\ particles from \nonrel\ ones
is an essential part of the NL problem, the energy range must
comfortably span $m_p c^2$ to be realistic.  If $\Emax=10$\,GeV is used,
with $u_0=5000$\,\kmps, and $\Eth= 10$\,MeV, equation~(\ref{eq:cell})
gives $\sim 1400$ and equation~(\ref{eq:step}) gives $\sim 4\xx{4}$.
Now, the computation may be possible, even with the 3-D requirement, but
the hybrid simulation can't fully investigate MFA since electron return
currents are not modeled.  The exact microscopic description of the
system is not currently feasible.

It's hard to make a comparison in run-time between PIC simulations
and the \MC\ technique used here since we are not aware of any published
results of 3-D PIC simulations of \nonrel\ shocks that follow particles
from fully \nonrel\ to fully \rel\ energies. A direct comparison of 1-D
hybrid and \MC\ codes was given in \citet{EGBS93} for energies
consistent with the acceleration of diffuse ions at the quasi-parallel
Earth bow shock. Three-dimensional hybrid PIC results for \nonrel\
shocks were presented in \citet{GE2000} and these were barely able to
show injection and acceleration given the computational limits at that
time. As for the \MC\ technique, all of the 16 nonlinear simulations
presented in this paper were completed in approximately 4 days on a
parallel computing cluster, employing around 30 processors; enough
statistical information was accumulated to restrict the uncertainty in
the self-consistent value of $\rtot$ to about 5 percent. 
Increasing the
dynamic range of the simulations to SNR-like energies ($\Emax
\approx 1$\,PeV) would require 2-4 times more computation time.
Thus, realistic \mc\ SNR models are possible with modest
computing resouces.

Despite these limitations, PIC simulations are the only way of \SCly\
modeling the plasma physics of collisionless shocks.
In particular, the injection of thermal particles in the large amplitude
waves and time varying structure of the subshock can only only be
determined with PIC simulations
\citep[e.g.,][]{NPS2008,Spitkovsky2008}. Injection is one of the most
important aspects of DSA and one 
where analytic and \mc\ techniques have large uncertainties.

\section{B. Analytic Precursor Approximation procedure}

\label{fastpush}

Here we describe our \FP\ (APA) where we introduce
thermal particles into the MC simulation,
not far upstream, but at some position $\xFP<0$ close
to the subshock. This
procedure has two purposes. First, it saves computational
time because we don't have to trace these particles
along the extended shock precursor. Second, we use the APA
to simulate the turbulence dissipation in the shock precursor:
with this procedure we incorporate the analytic description
of the heating of the thermal gas due to the turbulence dissipation
into the Monte Carlo model of particle transport.

In the next two subsections we
describe how the momentum and energy fluxes in shock precursor are
calculated in the absence of turbulence dissipation (part~\ref{noapa}),
and then explain how this scheme is changed when the \FP\ procedure is
invoked to model the thermal plasma heating by turbulence dissipation
(part~\ref{yesapa}).

\subsection{B.1. Precursor without Analytic Approximation}

\label{noapa}

\subsubsection{B.1.1. Inherent Quasi-Adiabatic Heating
 and Subshock Definition in Monte Carlo Simulation}

\label{mcadiabatic}

It is worth pointing out that even if the turbulence dissipation
is not included in our simulation, particles in the precursor
will still be weakly heated.
Just like an ideal collisionless gas put in a slowly shrinking volume
will heat up adiabatically due to elastic collisions of particles with
in-moving walls, the particles in our MC simulation traveling in a
compressing shock precursor will heat up according to the adiabatic
law $\Pth \propto \rho^{\gamma} \propto u^{-\gamma}$ due to
elastic scattering in the decelerating local plasma frame.
This will be true as long as the particles have enough time to adjust
to the changing flow speed. Quantitatively, the criterion
for adiabatic heating is
\begin{equation}
  \label{adiabaticcondition}
  \taucoll \ll \taucomp,
\end{equation}
where the collision time, $\taucoll$, is the ratio of the mean
free path to the particle speed, (for nonrelativistic particles
$\taucoll=m_p c/(e \Beff)$),
and the compression time $\taucomp$ is the temporal scale on which the
speed of plasma that the particle is traveling in changes significantly,
that is, by a value comparable to the particle speed $v$. Assuming $v
\ll u$, $\taucomp$ is approximately the ratio of the length on which the
flow speed changes by $v$ to the advection speed $u$:
$\taucomp=(v/|(du/dx)|) / u$. It also follows from the definition of
$\taucoll$ that the angular distribution of the particles remains
isotropic if (\ref{adiabaticcondition}) is true. The
condition~(\ref{adiabaticcondition}) is therefore
\begin{equation}
\left| \frac{du}{dx} \right| \ll \frac{v e \Beff}{u m_p c},
\end{equation}
restricting the flow speed gradient to small enough values, where
adiabatic compression operates.

If the magnitude of the flow speed gradient $|du/dx|$ is too large
and (\ref{adiabaticcondition}) doesn't hold, the increase
of particle energies will be faster than adiabatic, and the angular
distribution function will become non-isotropic, with more particles
moving against the gradient than along it in the plasma frame. One
obvious place where this occurs is the subshock.
In our simulation, the subshock in the non-linear,
self consistent solution gains a finite width, but the flow speed jump
at the subshock remains strong and rapid enough to efficiently
accelerate particles.  We define the point at which the
transition from the adiabatic to the non-adiabatic regime occurs,
$\xtr$, by the following condition:
\begin{equation}
  \taucoll(\xtr) = \frac13 \taucomp(\xtr),
\end{equation}
or, in terms of quantities available in the simulation,
\begin{equation}
\label{subshocklocation}
\left. \frac{du}{dx} \right|_{\xtr} =
- \frac13 \frac{v e \Beff(\xtr)}{u(\xtr) m_p c}
,
\end{equation}
where $v=\sqrt{2 k_B T(\xtr)/m_p}$,
in which case $|\xtr|$ is comparable to the local
convective mean free path of
a thermal particle.  The factor $1/3$ is chosen arbitrarily but our
results are insensitive to it.  Close to the shock, at $\xtr < x < 0$,
the flow speed drops so rapidly that the particles are heated in a
non-adiabatic fashion. We can think of the non-adiabatic region as
a subshock with a finite thickness $|\xtr|$, and it is then
reasonable to define the pre-subshock quantities (denoted with index
`1') as the values at $x=\xtr$: $u_1 = u(\xtr)$, $\rsub =
u(\xtr)/u(x>0)$, etc.
We note that Equation~\ref{subshocklocation} is only used to define
the position of the subshock an is not used in any calculations.

\subsubsection{B.1.2. Direct Calculation of Momentum and Energy Fluxes}

\label{calculatingfluxes}

The flux of momentum $\momentumflux(x)$ defined in~(\ref{mfasmoment}) is
used in our simulation to calculate the smoothing of the precursor
plasma flow $u(x)$, and the flux of energy $\energyflux(x)$ defined
in~(\ref{efasmoment}) is used to calculate the compression ratio $\rtot$
consistent with the particle acceleration. The moments of the particle
distribution function in~(\ref{mfasmoment}) and (\ref{efasmoment}) are
the components of the stress tensor. If the plasma heating by turbulence
dissipation is not modeled, and the \FP\ procedure is not
performed, then these moments are calculated in our simulation as
described below.  At select positions on the numerical grid spanning
from far upstream to some position downstream from the subshock
we sum the contributions of the particles that cross these positions to
calculate the following:
\begin{eqnarray}
\label{mfcalc}
\int p_x v_x f(x,\pvector)d^3p &=&
         \sum\limits_{i} p_\mathrm{i,\,x} v_\mathrm{i,\,x} w_i,\\
\label{efcalc}
\int K v_x f(x,\pvector)d^3p &=&
         \sum\limits_{i} K_i v_\mathrm{i,\,x} w_i.
\end{eqnarray}
Here the summation is taken over all particles crossing the position $x$
at which the moments are calculated. The index $i$ represents the
considered particle, $p_x$ ($v_x$) is the $x$-component of a
particle's of momentum (velocity),
$K$ is the kinetic energy, all measured in the shock frame, and $w$ is
the weight of the particle defined as
\begin{equation}
  w_i = \left| \frac{u_0}{v_\mathrm{i,\,x}} \right| \frac{n_0}{N_p}.
\end{equation}
In this definition the ratio $|u_0/{v_\mathrm{i,\,x}}|$ is the weighting
factor accounting for the fact that in our simulation particles crossing
the position $x$ at some angle to the flow do it less frequently than
particles crossing parallel to the flow, $n_0$ is the upstream number
density of the plasma and $N_p$ is the number of simulation particles.

If the particle distribution is isotropic in the local plasma
frame moving at speed $u(x)$ relative to the shock, then the quantities
calculated in (\ref{mfcalc}) and (\ref{efcalc}) can be expressed in the
following way:
\begin{eqnarray}
\label{momentumviapressures}
\int p_x v_x f(x,\pvector)d^3p &=&
\rho(x) u^2(x) + P_p(x), \\
\label{energyviapressures}
\int K v_x f(x,\pvector)d^3p &=&
\frac{1}{2} \rho(x) u^3(x) + w_p(x)u(x) \ ,
\end{eqnarray}
where $P_p(x)$ is the pressure and $w_p(x)$ is the enthalpy
of the particles.  In the
vicinity of the subshock, however, the isotropy
assumption breaks down (see the discussion in Appendix~\ref{mcadiabatic}
and the note on anisotropy in Appendix~\ref{special_th_vs_cr}), and the
concept of isotropic pressure is not applicable. The fact that we
directly calculate the moment of the distribution function
in~(\ref{mfasmoment}) by evaluating the sum~(\ref{mfcalc}), instead of
approximating the momentum flux with~(\ref{momentumviapressures}),
ensures that we properly account for the effects of the anisotropy of
particle distribution. This turns out to be important for
self-consistently determining the flow speed $u(x)$ near the subshock,
which controls the subshock compression and the subsequent particle
injection and acceleration efficiency.

When plasma heating by turbulence dissipation is modeled, we replace the
calculation~(\ref{mfasmoment}) with an analytic approximation assuming
isotropic particle pressure, but we take special care to be certain that
this approximation is only done far enough from the shock, where the
isotropy approximation is applicable. This procedure is described below.

\subsection{B.2. Modeling Precursor Heating with the \FPshort}

\label{yesapa}

\subsubsection{B.2.1. Particle Introduction Position}

The position at which the thermal particles are introduced must be close
enough to the shock so that the analytic description
of heating applies to most of the precursor extent, but far enough away
from the non-adiabatic region $\xtr < x < 0$, so that the analytic
approximation remains valid where applied.  In our simulation we chose
the particle introduction position, $\xFP$, so that the
condition~(\ref{adiabaticcondition}) is only marginally valid at this
position. We formalize it as
\begin{equation}
\taucoll(\xFP) = M \taucomp(\xFP),
\end{equation}
which is equivalent to
\begin{equation}
\label{xfastpush}
\left. \frac{du}{dx} \right|_{\xFP} = - M \frac{v e \Beff(\xFP)}{u(\xFP) m_p c}.
\end{equation}
where we chose $M = 0.1$ and $v = \sqrt{2 k_B T_0/m_p}$.  At $x<\xFP$ we
describe the thermal particle distribution function as a
Maxwellian with the temperature defined by (\ref{pressuregrowth}) and
(\ref{idealgaslaw}), and at $x>\xFP$ we use the Monte Carlo
simulation to describe the more complex particle dynamics.

\subsubsection{B.2.2. Momentum Space Distribution of Introduced Particles}

In order to include the effects of heating in the model, we must
introduce thermal particles at $\xFP$ as if they were heated in the
precursor, i.e., their temperature $T(\xFP)$ must be determined
by~(\ref{pressuregrowth}) and (\ref{idealgaslaw}). We therefore choose
the magnitude of every particle's momentum $p$ in the local plasma frame
distributed according to Maxwell-Boltzmann distribution with probability
density
\begin{equation}
  f(p) = \frac{4}{\sqrt{\pi}} \left( \frac{1}{2 m k_B T(\xFP)} \right)^{3/2}
  p^2 \exp{\left({-\frac{p^2}{2 m k_B T(\xFP)}}\right)}.
\end{equation}

The angular distribution of momenta of the
introduced particles is a major issue of concern in doing simulation
like ours because it determines the particle
injection rate. We are replacing the dynamics of particles at $x<\xFP$
with an analytical description, consequently we must distribute
particles in $p$-space at $\xFP$ the way they would be distributed
having traveled from far upstream and reaching $\xFP$ {\it for the first
time}. This is equivalent to calculating a $p$-space distribution of
particles incident on a {\it fully absorbing boundary} at $\xFP$ after
scattering in a non-uniform flow $u(x)$. This is easy to do analytically
if all particles have a plasma frame speed $v$ less than the flow speed
$u(\xFP)$ (because then all particles crossing position $\xFP$ do it for
the first and the last time), and fairly complicated otherwise. We
assume $v<u(\xFP)$ in further reasoning, which is justified by the fact
that we find $M_s(\xFP) \gtrsim 3$ in most cases.

As was stated earlier, we assume that the angular distribution of
momenta of the introduced thermal particles is isotropic in the plasma
frame.  When these particles cross a position fixed in the shock
frame, their flux must be `weighted' to account for the fact that the
number of particles arriving at $\xFP$ in a unit time is proportional to
the cosine of the angle that their shock frame velocity
$\vvector_\mathrm{sf}$ makes with the $x$-axis.
This can be done by assuming a probability density of
$\mu = v_x/v$ ($v$ is the magnitude of the particle plasma frame
velocity and $v_x$ its $x$-component) as
\begin{equation}
\label{fluxweighting}
f(\mu)=\frac12 \left(1 + \mu \frac{v}{u}\right),
\end{equation}
where $u=u(\xFP)$. It is normalized so that the probability
$Pr(\mu_0 < \mu < \mu_0 + d\mu_0) = f(\mu_0) d\mu_0$, and the functional
form of (\ref{fluxweighting}) comes from the assumption that $f(\mu)
\propto v_\mathrm{sf,\:x}=u + \mu v$.

\subsubsection{B.2.3. Heating of the Upstream Plasma}

After the thermal particles are introduced at $\xFP$ and start to
  propagate in the shocked flow, we have to
  calculate the momentum and energy fluxes throughout the shock, for use
  in our iterative procedure. Because we didn't propagate the thermal
  particles at $x<\xFP$, and because we must model the momentum flux
  redistribution between the turbulence and the thermal particles due to
  heating, we calculate the corresponding moments of particle
  distribution function the following way:
\begin{eqnarray}
\label{apamomentum}
\int p_x v_x f(x,\pvector)d^3p &=&
\left\{
  \begin{array}{ll}
    \sum\limits_{all\:i} p_\mathrm{i,\,x} v_\mathrm{i,\,x} w_i, & \mathrm{if} \; x > \xFP, \\
    \rho(x) u^2(x) + \Pth(x) +
    \sum\limits_{i \in CR} p_\mathrm{i,\,x} v_\mathrm{i,\,x} w_i, & \mathrm{if} \; x < \xFP,
  \end{array}
\right. \\
\label{apaenergy}
\int K v_x f(x,\pvector)d^3p &=&
\left\{
  \begin{array}{ll}
    \sum\limits_{all\:i} K_i v_\mathrm{i,\,x} w_i, & \mathrm{if} \; x > \xFP, \\
    \displaystyle\frac12 \rho(x) u^3(x) + \displaystyle\frac{\gamma}{\gamma-1}\Pth(x)u(x) +
    \quad & \\ \qquad \qquad \qquad \qquad \qquad
      + \sum\limits_{i \in CR} K_i v_\mathrm{i,\,x} w_i, & \mathrm{if} \; x < \xFP.
  \end{array}
\right.
\end{eqnarray}
The summation for $x>\xFP$ is taken over all the particles crossing position
$x$, and for $x<\xFP$ the summation index $i\in CR$ only includes the CR (i.e., injected)
particles, while the contribution of the thermal particles is replaced by the
analytic approximation of this contribution. The thermal pressure in this approximation
is taken from the solution of~(\ref{pressuregrowth}).
For $\heatpar=0$ the equations~(\ref{apamomentum})
and (\ref{apaenergy}) produce the same results
(within intrinsic random deviations of the Monte Carlo code)
as the calculation~(\ref{mfcalc}) and (\ref{efcalc}).

In the region $\xFP < x < 0$ the heating due to $\Lbar$ is ignored.
This may, in principle,
underestimate the heating of the upstream gas, but we find that this
is a negligible effect. We prove it by running a simulation with
another $\xFP$, even farther away from the subshock, and making sure
that the results are not affected significantly.

\section{C. Note on Differentiation between Thermal and CR particles}

\label{special_th_vs_cr}

As was mentioned earlier, we call a particle a thermal particle or a CR
one based on its history: a CR particle is one that has gotten injected
into the acceleration process by having crossed the shock from the
downstream to the upstream region at least once\footnote{Although
the shock gains a finite width in the Monte Carlo simulation, we
define the backward shock crossing as moving from $x>0$ to $x<0$.}.
This criterion is used
at two places in the model.  First, it is used to calculate the spectrum
of pressure driving the CR streaming instability $\Pcrhat(x,p)$: only
the contribution of injected particles is included in
$\Pcrhat(x,p)$.  Second, when we calculate the thermal particle
pressure $\Pth(x)$ at $x<\xFP$ using (\ref{dampingparameterization})
and (\ref{pressuregrowth}), and then introduce thermally distributed
particles at $x=\xFP$ and continue the calculation
of pressure for $x>\xFP$ from their trajectories as they elastically
scatter in the flow, we implicitly
assume that the dissipated energy of the turbulence goes directly
into the thermal energy of the particles that have not been injected,
i.e., thermal particles in our definition.

If a description of particle-wave interactions in strong turbulence
existed that explicitly described how particles of different energies
participated differently in the instability generation and the
turbulence dissipation, a criterion like ours would not be
needed. However, such a description is not available and we believe that our way
of separating thermal and superthermal (CR) particles for purposes of
describing the instability growth and the turbulence dissipation grasps
the essential non-linear effects in the shock structure.
PIC simulations are, in principle, able to tackle this problem
exactly but, as we mentioned earlier, they are extremely computationally
expensive.

We would like to point out an important consequence of our using the
`thermal leakage' model of particle injection into
the acceleration process.  Our simulation follows histories of charged
particles from their `childhood', when their speeds in the plasma frame
are small compared to the bulk flow speed, to `maturity', when they
become relativistic. Unlike most semi-analytic descriptions of DSA, our
model doesn't skip the `adolescence' stage of particles, when after one
or a few shock crossings the particles have speeds comparable to or
slightly greater than the bulk flow speed. In the absence of these
`adolescent' particles, it is typically assumed that the jump in only
the thermal particle pressure across the subshock determines the
strength of the latter, and the superthermal part of the particle
spectrum is continuous at the subshock and does not influence it.  But
the `adolescent' particles that the Monte Carlo model does describe are
not energetic enough to be insensitive to the subshock, and at the same
time they have a strong anisotropy in the plasma frame, and therefore do
not obey the Rankine-Hugoniot relations. This modifies the conservation
laws across the subshock, because in the total kinetic pressure
$P_p(x)=\Pth(x) + \Pcr(x)$ the term $\Pcr(x)$ is not continuous at the
subshock due to the contribution of the intermediate energy particles
that it contains.

\section{D. Compression ratio, Turbulence and Escaping Particles}

\label{compressionratio}

Equation (10) in \citet{EMP90} relates the fraction of energy flux
carried away by escaping particles $\qesc$ to the total shock
compression ratio, $\rtot$. It assumes no
magnetic field amplification and a polytropic index of downstream gas
equal to $5/3$ (in other words, neglects the effect of relativistic
particles on the overall compressibility of the gas). In order to search
for a $\rtot$ consistent with the shock structure and with
particle escape, we derive a similar relationship that
would account for the presence of magnetic turbulence and for the
contribution of the relativistic particles. The problem is complicated
by having particles of intermediate (mildly relativistic) energies and
by the value of the magnetic field dependent on the particle
acceleration.

Writing equations (\ref{fluxmass}), (\ref{fluxmomentum}) and
(\ref{fluxenergy}) for a point downstream of the shock, sufficiently
far from it, so that the distribution of particle momenta is isotropic,
and the approximations~(\ref{momentumviapressures}) and
(\ref{energyviapressures}) are valid, and
denoting the corresponding quantities by index `2', we get:
\begin{eqnarray}
\label{cons1}
\rho_2 u_2 &=& \rho_0 u_0, \\
\label{cons2}
\rho_2 u_2^2 + \Pptwo + \Pwtwo &=& \rho_0 u_0^2 + \Ppzero + \Pwzero\equiv\upstreammomentumflux, \\
\label{cons3}
\frac12 \rho_2 u_2^3 + \wptwo u_2 + \Fwtwo + \Qesc&=&
     \frac12 \rho_0 u_0^3 + \wpzero u_0 + \Fwzero\equiv\upstreamenergyflux.
\end{eqnarray}
The particle gas enthalpy $w_p$ is $w_p=\epsilon_p + P_p$, and the
internal energy $\epsilon_p$ of gas is proportional to
the pressure $P_p$. Introducing the quantity $\gammabar$ so that
$\epsilon_p = P_p/(\gammabar-1)$, one can write
\begin{equation}
\label{fppp}
w_p u=\frac{\gammabar}{\gammabar-1}P_p u
\end{equation}
The value of $\gammabar$ is averaged over the whole particle
spectrum, and it ranges between $5/3$ for a nonrelativistic
and $4/3$ for an ultra-relativistic gas.
The local value of $\gammabar$ can be easily calculated in our
code from the particle distribution, along with
$P_p$ and $\epsilon_p$, as $\gammabar=1 + P_p / \epsilon_p$.
Similarly, one can define $\deltabar=F_w/(u P_w)$ and calculate a local
value of $\deltabar$ anywhere in the code in order to express
\begin{equation}
\label{fwpw}
F_w=\deltabar \cdot P_w u \ .
\end{equation}
For $V_G \ll u$ and \Alf ic turbulence, one expects $\deltabar
\approx 3$, [see eq.~(\ref{fwdef})].

Substituting (\ref{fppp}) and (\ref{fwpw}) into the above equations
and introducing $\rtot=u_0 / u_2$, we can eliminate $\rho_2$ using
(\ref{cons1}) and $\Pptwo$ using (\ref{cons2}), which allows us to
express from (\ref{cons3}) the quantity $\qesc \equiv \Qesc /
\upstreamenergyflux$ as
\begin{equation}
\label{qescrtot}
\qesc = 1 + \frac{A/\rtot^2 - B/\rtot}{C},
\end{equation}
where
\begin{eqnarray}
A&=& \frac{\gammabar_2 + 1}{\gammabar_2 - 1} ,\\
B&=& \frac{2 \gammabar_2}{\gammabar_2 - 1}
     \left( 1 + \frac{\Ppzero + \Pwzero - \Pwtwo}{\rho_0 u_0^2}\right) +
            \frac{2 \deltabar_2 \Pwtwo}{ \rho_0 u_0^2 },\\
C&=& 1 + \frac{2 \gammabar_0}{\gammabar_0 - 1}\frac{\Ppzero}{\rho_0 u_0^2} +
            \frac{2 \deltabar_0 \Pwzero}{\rho_0 u_0^2}.
\end{eqnarray}
Note that $\rho_0 u_0^2 / \Ppzero = \gammabar_0 M_s^2$, where
$\gammabar_0=\gamma=5/3$ due to the absence of CRs far upstream,
and, because we assume a seed turbulence far upstream, that
provides a Bohm regime of scattering to all particles,
one can write that far upstream $\Delta B \approx B_0$, making
$\rho_0 u_0^2 / \Pwzero=2 M_A^2$.

The quantity $\qesc$ is readily available in the simulation after
the end of any iteration.
Comparing it to the value predicted by (\ref{qescrtot}), we
evaluate the self-consistency of the solution and make the correction to
$\rtot$, if necessary, for further iterations. For making these
corrections it is helpful to use in the simulation the inverse of
(\ref{qescrtot}), the physically relevant branch of which is
\begin{equation}
\label{rtotqesc}
\rtot = \frac{2 A}{B - \sqrt{B^2 - 4 A C (1-\qesc)}}.
\end{equation}

It is important to emphasise here that an iterative procedure
similar to~(\ref{iteration}) is still
required to find the compression ratio $\rtot$ of a non-linearly
modified shock, because quantities $\qesc$, $\Pwtwo$ and $\gammabar_2$
depend on $\rtot$, so (\ref{rtotqesc}) only provides a practical
way to perform the iterations.

\section{E. Note on Subshock Properties in the Presence of MFA}

\citet{KJG2002} showed, using a thermal leakage injection model
similar to what is used here, that in a non-linearly modified shock with
Bohm diffusion and a sonic Mach number $\Mszero$, the subshock sonic
Mach number scales as $\Msone \sim 2.9 \Mszero^{0.13}$ with the
corresponding subshock compression ratio $\rsub =
4/(1+3/\Msone^2)$. Numerically, for $\Mszero=400$ it gives $\Msone
\approx 6.3$ and $\rsub \approx 3.7$.

The model of \citet{KJG2002} does not include magnetic field
amplification and, as we have shown, in
our case the values of the subshock sonic Mach number and compression
ratio are quite different.  For instance, for our model with MFA, but
without magnetic turbulence damping ($\heatpar=0$) we have found that
$\Mszero = 426$ results in $\Msone \approx 44$
and $\rsub \approx 3$. The
value of $\Msone$ in the absence of precursor plasma heating is
determined by the high CR pressure $\Pcr(x)$, and the value of $\rsub$
in the high Mach number shock with MFA is largely controlled by the
pressure of the amplified magnetic turbulence $P_w(x)$ rather than by
the thermal pressure $\Pth(x)$. The situation is changed when the
turbulence dissipation operates, because it dampens $P_w(x)$ and
increases $\Pth(x)$, which reduces $\Msone$.

Our getting such a high value of $\Msone$ is important because,
just like \citet{KJG2002}, we have found in
Section~\ref{injectionvsdissipation} that $\Msone \approx 10$ is a
`breaking point' in the thermal leakage model of particle injection: the
injection rate depends weakly on $\Msone$ when $\Msone \gtrsim 10$, but
increases rapidly with decreasing $\Msone$ if $\Msone < 10$.

%\clearpage
\begin{figure}
\epsscale{.50} \plotone{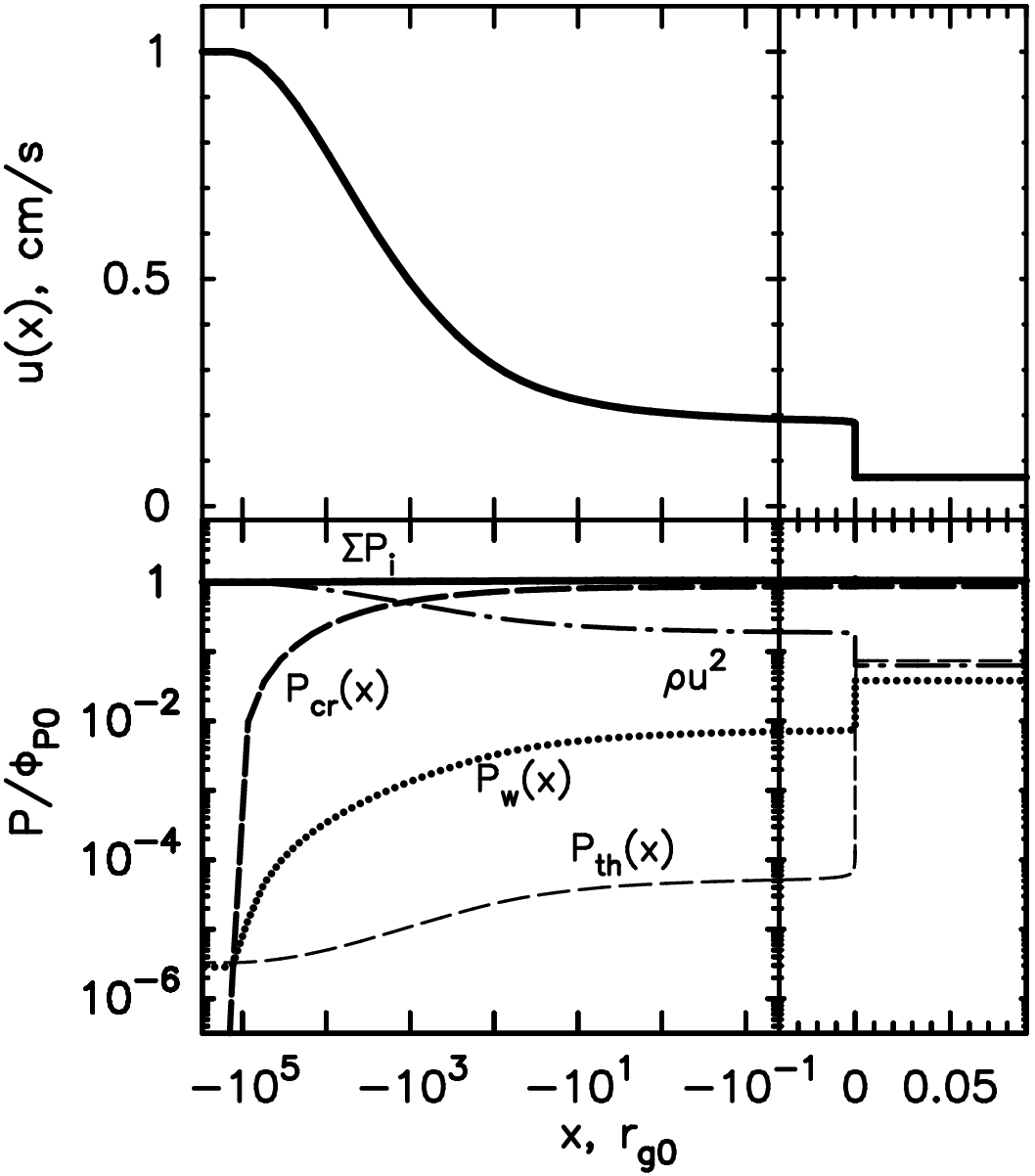}
\caption{Illustration of momentum flux balance in the non-linear
  simulation with $\heatpar=0$ in the \coldismNoT\ case.  The thin
  dashed line is the thermal pressure $\Pth(x)$, the thick dashed line -
  the CR pressure $\Pcr(x)$, the dotted line is the magnetic turbulence
  pressure $P_w(x)$, the dash-dotted line is the dynamic pressure
  $\rho(x) u^2(x)$, and the thick solid line is the sum of the four, the
  total momentum flux. All quantities are normalized to the far upstream
  value of the total momentum flux.
  \label{fig_momflx}}
\end{figure}
%\clearpage

As an illustration for the above discussion, we
show in Figure \ref{fig_momflx} the various constituents of the momentum
flux across the shock with $\heatpar=0$ in the \coldism\ case.
It's clear that upstream of the shock the dominant contributor to the
momentum flux is the CR pressure $\Pcr(x)$ and, therefore, the precursor
compression is mainly determined by $\Pcr(x)$. For this particular set
of parameters, $\Pcr(x)$ results in a decrease in the flow speed $u(x)$
by a factor of $\sim 5.4$ from its upstream value $u_0$ to the
pre-subshock value $u_1$.
The temperature in the precursor only increases adiabatically for
$\heatpar=0$ and this results in an increase over the far upstream
temperature $T_1/T_0 \approx 3.3$, thus reducing the local sonic Mach
number to $\Msone \approx 44$.
%The thermal pressure and the magnetic pressure remain low
%for $x<0$ and don't come into play in determining the value of $\Msone$.

The subshock compression ratio $\rsub$ is determined by the change in
the different constituents of the momentum flux across the subshock.
$\Pcr(x)$, although large, changes little across the
subshock\footnote{
$\Pcr(x)$ is discontinuous at $x=0$ in our simulation
due to the contribution to it of the particles that have crossed the
shock only a few times. The jump in $\Pcr(x)$ in this case is small
compared to the jump in thermal and magnetic pressures across the
subshock, but can be significant for $\heatpar>0$, as seen in
Figure~\ref{fig_partpress}.},
and what determines $\rsub$ is the change in $\Pth(x)$ and $P_w(x)$. The
latter, as the plots in Figure \ref{fig_momflx} show, contributes
significantly to the momentum flux.  This is an important point because
the values of $\Befftwo$ and $P_w(x)$ depend on $\rsub$ and $\rtot$ in a
non-linear way making the traditional Rankine-Hugoniot relations
inapplicable for determining $\rsub$.
Relation (\ref{rtotqesc}) can be used to
iteratively calculate the resulting compression ratio $\rtot$, and it
results in $\rtot=16$ and $\rsub=2.95$, as Table \ref{sumnl_cold}
shows. As we can see, the effect of magnetic turbulence pressure tends to decrease
$\rsub$ compared to the case $P_w(x) \ll \Pth(x)$ (as in the latter case
one would expect $\rsub\approx 4$ for $\Msone \approx 44$).

\clearpage

\bibliographystyle{apj}
%%\bibliography{c:/a_a_TOP/bibTeX/bib_DCE}
\bibliography{bib_DCE}

\end{document}